\documentclass[%
 preprint, 
 amsmath,amssymb,
 aps, physrev,
]{revtex4-2}

\usepackage{graphicx}
\usepackage{dcolumn}
\usepackage{bm}
\usepackage{overpic}
\usepackage{xcolor}
\usepackage[normalem]{ulem}
\usepackage{makecell}
\usepackage{hyperref}

\newcommand{\dimhead}{d_k}
\newcommand{\dimstate}{d_{\text{model}}}
\newcommand{\numhead}{n_{\text{head}}}
\newcommand{\dimobs}{d_{\text{obs}}}
\newcommand{\amass}{\text{am}}
\newcommand{\gust}{\text{g}}
\newcommand{\resp}{\text{c}}

\newcommand{\revision}[1]{\textcolor{black}{#1}}

\begin{document}

\title{\textbf{Attention on flow control: transformer-based reinforcement learning for lift regulation in highly disturbed flows} 
}%

\author{Zhecheng Liu}%
\email{Contact author: zliu163@ucla.edu}
\author{Jeff D. Eldredge}
\affiliation{%
 Mechanical and Aerospace Engineering, University of California, Los Angeles, CA 90095-1597, USA
}%

\date{\today}

\begin{abstract}
A linear flow control strategy designed for weak disturbances may not remain effective in sequences of strong disturbances due to nonlinear interactions, but it is sensible to leverage it for developing a better strategy. In the present study, we propose a transformer-based reinforcement learning (RL) framework to learn an effective control strategy for regulating aerodynamic lift in \textcolor{black}{arbitrarily long} gust sequences via pitch control. \textcolor{black}{The random gusts produce intermittent, high-variance flows observed only through limited surface pressure sensors, making this control problem inherently challenging compared to stationary flows.} The transformer addresses the challenge of partial observability from the limited surface pressures. We demonstrate that the training can be accelerated with two techniques---pretraining with an expert policy (here, linear control) and task-level transfer learning (here, extending a policy trained on isolated gusts to multiple gusts). We show that the learned strategy outperforms the best proportional control, with the performance gap widening as the number of gusts increases. The control strategy learned in an environment with a small number of successive gusts is shown to effectively generalize to an environment with an arbitrarily long sequence of gusts. We investigate the pivot configuration and show that quarter-chord pitching control can achieve superior lift regulation with substantially less control effort compared to mid-chord pitching control. Through a decomposition of the lift, we attribute this advantage to the dominant added-mass contribution accessible via quarter-chord pitching.
\end{abstract}

\maketitle
\section{Introduction}
\label{sec:introduction}
An interest in large manned air vehicles has been extended to encompass unmanned air vehicles (UAVs) in recent years, particularly to serve roles in intelligence, surveillance, and delivery \cite{LAGHARI20238}. When performing a wide range of operations, it is expected for UAVs to achieve stable flight even in extreme environments where they may encounter sizable atmospheric disturbances (gusts) \cite{Fukami2023}. However, a large force and moment transient experienced during gust encounters may impact the flight path and safety of the aircraft and, over time, cause structural fatigue to the vehicle \cite{Wu2019}. Therefore, many recent studies have been devoted to investigating the underlying unsteady aerodynamic physics of gust encounters \cite{Eldredge_Jones_2019, Jones2022, Narayanan2024} and designing control strategies to mitigate gust effect \cite{Oduyela2014, Sedky_2022, Kerstens2011}. 

Each combination of the control objectives (e.g., separation prevention, lift enhancement, or drag reduction), actuation methods (e.g., pitching, control surface, or synthetic jets), and the availability of measurements (e.g., pressure, velocity, acceleration, or force and moment), presents unique challenges and requires a tailored design approach. The control problem is manageable when gusts are of small amplitude, so that one can rely on linear control theory designed on a linear model of the aerodynamics. For example, Sedky et al. \cite{Sedky_2022} designed a controller to mitigate the lift variation on an airfoil by modeling the lift response to pitching kinematics and transverse gust from the superposition of the classical K\"ussner model \cite{Kussner1932} and Wagner model \cite{Wagner1925}. Brunton and Rowley \cite{BRUNTON2013} developed a empirical state-space form of Theodorsen model \cite{Theodorsen1949} and leveraged it to design a robust controller to track a reference lift profile with $H_\infty$ loop-shaping approach. However, the use of classical aerodynamic models relies strictly on the fully attached flow assumption, and linear descriptions of the aerodynamics tend to fail if flow separation is induced, as is prone to occur when the disturbance is of large amplitude. In such circumstances one can no longer rely on superposition of indicial disturbance responses from sequences of gusts. Linear control tends to show degraded performance or even fail in these strong-gust conditions. In the absence of linear superposition, the control task is further confounded by the vast parameter space of the gusts and the strong nonlinear interactions with the wings \cite{Jones2020, Jones2022}. 

Therefore, this challenge has spurred a large number of studies on modeling the nonlinear flow response and designing the appropriate control strategies when the flow is not fully attached. One popular approach is to linearize the flow around an equilibrium condition \cite{Rowley2005, Ma2011, Brunton_Rowley_Williams_2013}. However, the limitation of this approach is that the model would be accurate only when the flow condition is near the equilibrium condition. Therefore, a gain-scheduling approach can be used to construct a series of linear models based on a series of equilibrium points. For example, Dawson et al. \cite{Dawson2015} constructed $8$ linear models for each corresponding to an incremental range of angle of attack, and these models could be switched during random pitch motion. However, they found that switching between the models leads to some degradation of accuracy when the pitching rate becomes relatively large. To address this problem, some other approaches have been pursued that incorporate proxy parameters of the relevant fluid flow physics into the system equations, such as linear parameter-varying model \cite{Hemati2017}, Goman-Khrabrov model \cite{Goman1994, Williams2018}, and the time delay and decay model \cite{An2016}. Although these models show improved performance, their effectiveness depends heavily on the appropriate selection and fine-tuning of proxy parameters, which raises concerns about their generalizability in the vast parameter space of gust disturbances \cite{Jones2020} and the diversity of available actuation mechanisms. Therefore, the development of a more generalizable and adaptive control strategy is still an open and compelling challenge.

Fortunately, the development of machine learning and data science potentially provides us with data-driven tools to better model and control these complex flow problems. Among them, reinforcement learning (RL)---particularly when outfitted with a deep network for nonlinear function approximation, commonly referred to as deep reinforcement learning---has demonstrated notable success in a variety of flow control tasks. The examples include exploring effective swimming strategies \cite{Novati2017, Verma2018}, learning an optimized foil flapping motion \cite{Wang2024}, reducing drag of a bluff body \cite{Fan2020, Rabault2019,Xia2024,YAN2025}, mitigating lift variation of an airfoil \cite{Liu2025, Beckers2024, Renn2022}, and many others.

\textcolor{black}{However, almost all of the studies above address control of a statistically stationary flow. In particular, in most of them the flow behaves periodically without any actuation, so the RL agent is free to explore a single waveform (e.g., cylinder rotation, synthetic-jet duty cycle, or flapping stroke) that maximizes the reward. In contrast, in the present case, the gust environment tackled is not stationary at all: every random vortex that sweeps over the wing resets the flow to a new, a priori unknown state, and generates a significantly different aerodynamic response even without actuation. Thus, from a RL point of view, the trajectories produced under the same control policy will span different regions of the state space, and their consequent cumulative rewards can differ by several orders of magnitudes. This high-variance property makes the policy and value networks prone to overfitting to whichever subset of trajectories is currently dominant in the training batch, which slows convergence. Additionally, in most of the previous work, the policy is evaluated in exactly the same statistical regime in which it is trained. Here, we demand the evaluation outside the training regime: the policy learned in a finite number of gusts needs to generalize well to an infinite number of gusts. These requirements place the flow control in a highly disturbed environment beyond the difficulty of canonical problems such as drag reduction or path planning in a statistically stationary flow.}

\textcolor{black}{Previous attempts to mitigate highly disturbed flows with RL made important progress, but each has left key gaps. In a flow context similar to the present paper, Beckers and Eldredge \cite{Beckers2024} trained a model-free agent in viscous, two-dimensional gust encounters for lift regulation of an airfoil using sparse surface pressures. Their policy and value networks were multi-layer perceptrons \revision{(MLP)} that processed a fixed history of observations. Such feed-forward models do not intrinsically model temporal dependencies, so the policy learned in a sequence of two gusts struggled when it was applied to long sequences of gusts. Furthermore, the learning curve converged at $5000$ episodes, which is prohibitively expensive from the perspective of CFD simulations. Therefore, for the purpose of reducing training cost, Liu et al.~\cite{Liu2025} proposed a model-based approach that learns a reduced-order surrogate model and optimizes the policy within the reduced space, which saved training cost significantly but must be refit when the disturbance class changes (e.g., the number of gusts), hence limiting generalization.}

\textcolor{black}{We address both limitations in this work. We make use of the machine learning concept of {\em self-attention} and adopt the transformer \cite{Vaswani2017}, the most widely-used machine learning architecture that exploits this concept to model temporal dependencies, identifies correlations between different (and possibly widely separated) parts of the input observation sequence to approximate an output. We use the transformer output to approximate a belief state, which is used in turn to choose a corresponding control action via the proximal policy optimization (PPO) algorithm \cite{schulman2017}. By this approach, we demonstrate that the policy learned in a small sequence of gusts generalizes well in long sequences. Although the transformer has been successfully applied to some fluid dynamics problems in recent years \cite{Wang2024,Weissenbacher2025}, we note that, prior to this study, it was still unknown if it could be applied to the highly-disturbed flows, due to the unique challenges we discussed above. Furthermore, to reduce the computational cost, we leverage the techniques of pretraining and task-level transfer learning. The pretraining relies on the expert data from a well-designed proportional control (P control) strategy. We find that the RL control strategy that emerges outperforms the P control baseline, with an increasingly pronounced advantage as the number of gusts grows. The task-level transfer learning takes the RL agent trained in a single-gust environment as the starting point for the training in a multi-gust environment. We demonstrate that both these so-called warm-start techniques significantly speed up convergence compared to training from scratch.}

\textcolor{black}{Additionally, both the previous works \cite{Liu2025,Beckers2024} only explored a mid-chord pitching configuration. To investigate how the location of the pivot point affects the control performance of the RL agent, we revisit the pivot configuration and find that a quarter-chord pitching strategy spends much less control effort while achieving the lift regulation better than in mid-chord pitching. By analyzing the vorticity field, pressure distribution, lift history, and the corresponding pitching history, we offer physical insights into the superiority of quarter-chord pitching control.}

\textcolor{black}{The remainder of the paper is organized as follows. The set-up and description of the flow control problem, and the details about the transformer-based RL framework, are discussed in Section \ref{sec:methodology}. Then, we analyze the performance of the RL control strategy across different scenarios and pivot locations, accompanied by a discussion of the underlying physical mechanisms, in Section \ref{sec:resultsanddiscussion}. Finally, we summarize and discuss the results in Section \ref{sec:conclusions}.}

\section{Methodology}
\label{sec:methodology}
\subsection{Flow control problem description}
\label{subsection:control_problem}
\begin{figure}[htbp]
    \centering
    \includegraphics[width=0.8\linewidth]{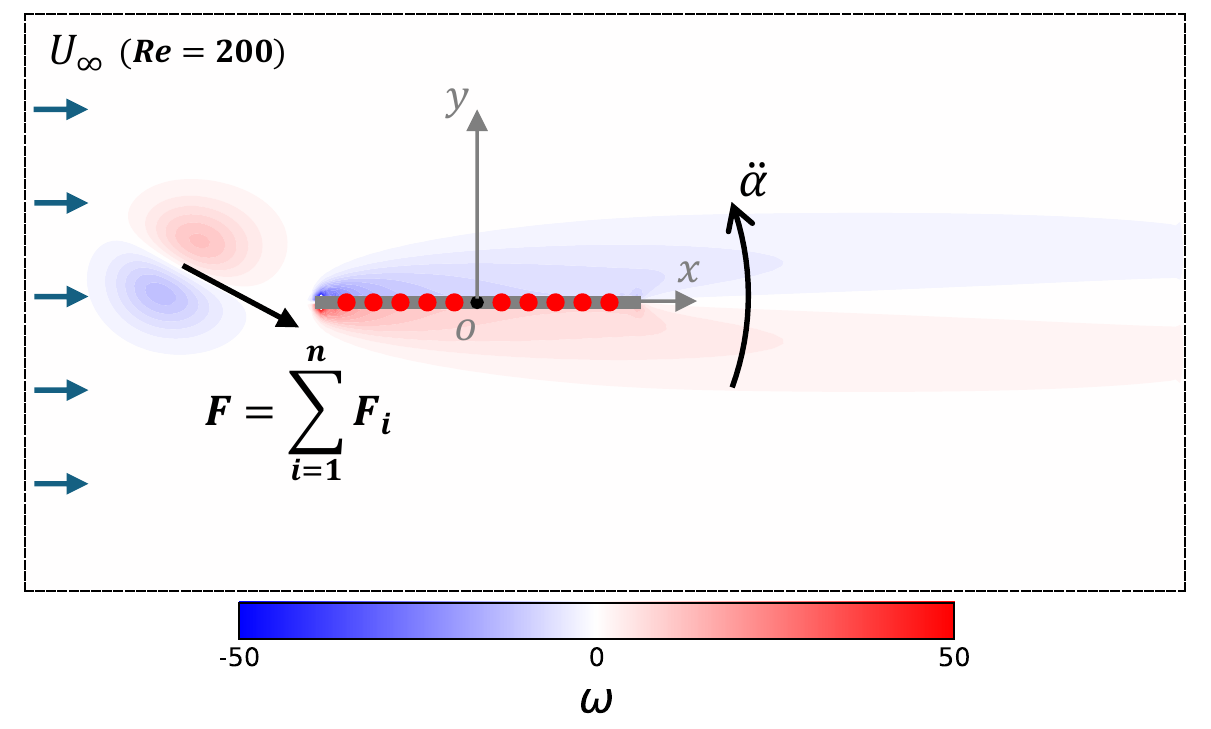}
    \caption{Control problem description. We only show the pivot point located at the mid chord for simplicity, shown as grey dot, but a quarter chord pivot point configuration is also investigated in this study. The pressure sensors are denoted as the red dots. The colorbar shown here is used throughout this study.}
    \label{fig:control_problem}
\end{figure}

In this study, we investigate a control strategy for regulating the lift of a flat plate of chord length $c$ in a two-dimensional flow with $Re=\rho U_{\infty} c/\mu=200$, where $U_{\infty}$ is the freestream velocity and $\rho$ and $\mu$ are the density and dynamic viscosity of the fluid, respectively. Specifically, the objective of the lift regulation in this study is to track a reference lift of zero during gust encounters by pitching the plate about a pivot axis. However, the proposed methodology is general and, within physical limits, can be easily extended to track arbitrary reference lift profiles with other actuation methods. As depicted in Fig. \ref{fig:control_problem}, gusts are introduced in the region upstream of the plate after the undisturbed flow is fully developed. For reference, the coordinate system is centered at the mid chord point of the plate when the plate is at zero pitch angle, with $x$ in the freestream direction.

Specifically, gusts are introduced successively, with each gust $i$ generated via a body force applied on the fluid, Gaussian-distributed in both space and time according to
\begin{equation}\label{eqn:Gaussian}
    \boldsymbol{F}_i(x,y,t) = \rho U_{\infty} c^2 \frac{(D_x,D_y)}{\pi^{3/2}\sigma_x \sigma_y \sigma_t} \exp\left[-\frac{(x-x_0)^2}{\sigma_x^2}-\frac{(y-y_0)^2}{\sigma_y^2}-\frac{(t-t_0)^2}{\sigma_t^2}\right],
\end{equation}
where $D_x$ and $D_y$ denote the dimensionless amplitudes of the Gaussian forcing in the respective coordinate directions, and $\sigma_x$, $\sigma_y$ specify the respective spatial spreads in these directions. The spatial center of the forcing is described by $x_0$ and $y_0$, while $t_0$ and $\sigma_t$ represent the temporal center and spread, respectively. The forcing field acts as a source term in the Navier–Stokes equations and thus generates a distributed region of vorticity in the fluid that subsequently convects and diffuses with the flow. The Gaussian spatial and temporal distributions ensure that this new vorticity comes in the form of vortices that are localized (but continuous) in space and time. This approach enables us to introduce gusts continuously, rather than creating them impulsively at a single instant, e.g., through initial conditions.

\begin{figure}[htbp]
    \centering
    \includegraphics[width=1.0\linewidth]{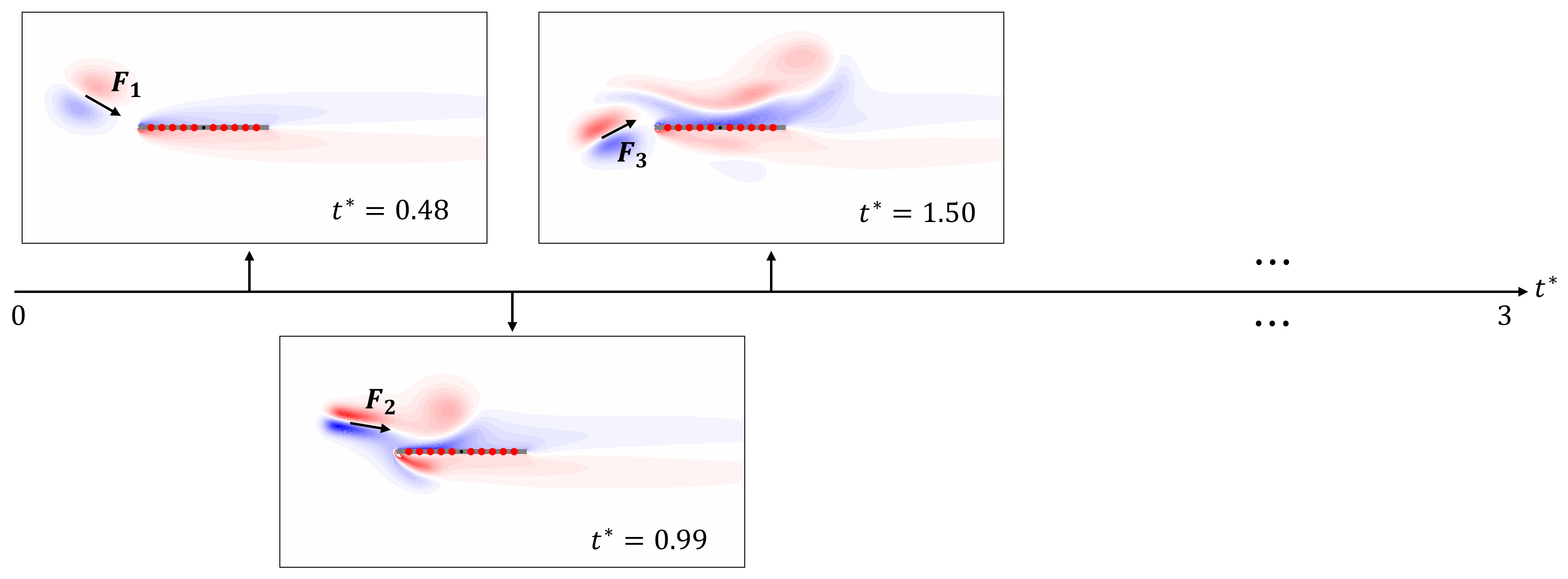}
    \caption{An example of environment with three successive gust under no control. $t^*$ is the convective time. The specific gust parameters, for this example, by the order of $(D_y,y_0/c,\sigma/c,U_\infty t_0/c)$ is: $(-1.0,0.25,0.1,0.5)$, $(-0.5,0.25,0.05,1.0)$, $(1.0,-0.1,0.075,1.5)$.}
    \label{fig:example_threegust}
\end{figure}

For training and testing purposes, several of the gust parameters in \eqref{eqn:Gaussian} vary from case to case, and, in the case of successive gusts, from gust to gust. The parameter space of \textit{n}-successive gusts (up to the maximum used in this study, eight) is given in Table \ref{tab:parameter_gust}. Some parameters remain fixed, others are chosen from a continuous range, and the temporal center of each gust is selected from a discrete set. The purpose of the time gap between the temporal centers of successive gusts is to simulate sequential gust encounters, as commonly observed in real disturbed flow environments such as wake regions. An example with three successive gusts is shown in Fig. \ref{fig:example_threegust} to illustrate how \textit{n}-successive gusts are implemented in this study.
\begin{table}
\caption{Parameter space of \textit{n}-successive gusts in this study.}
\label{tab:parameter_gust}
\begin{ruledtabular}
    \begin{tabular}{cccccccc}
      &  $D_x$ &  $D_y$ & $x_0/c$ & $y_0/c$ & $\sigma/c = \sigma_x/c = \sigma_y/c$ & $U_\infty t_0/c$ & $U_\infty\sigma_t/c$ \\
      \hline
      $1\text{st}$ \text{gust} & $1.2$   & $[-1, 1]$ & $-1.0$   & $[-0.25, 0.25]$ & $[0.05, 0.1]$ & $0.5$ & $0.2$ \\
      $2\text{nd}$ \text{gust} & $1.2$   & $[-1, 1]$ & $-1.0$  & $[-0.25, 0.25]$ & $[0.05, 0.1]$ & $\{0.9, 0.95, 1.0, 1.05, 1.1\}$ & $0.2$ \\
      $3\text{rd}$ \text{gust} & $1.2$  & $[-1, 1]$ & $-1.0$  & $[-0.25, 0.25]$ & $[0.05, 0.1]$ & $\{1.4, 1.45, 1.5, 1.55, 1.6\}$ & $0.2$ \\
      \textellipsis & \multicolumn{7}{c}{$\cdots$} \\
      $8\text{th}$ \text{gust} & $1.2$  & $[-1, 1]$ & $-1.0$  & $[-0.25, 0.25]$ & $[0.05, 0.1]$ & $\{3.9, 3.95, 4.0, 4.05, 4.1\}$ & $0.2$ \\
    \end{tabular}
\end{ruledtabular}  
\end{table}

The control input in this study is the angular acceleration $\ddot{\alpha}$ of the plate about a fixed pivot point, located at either mid chord or quarter chord. In order to simulate a practical control problem with limited bandwidth, we restrict the angular acceleration $\ddot{\alpha}$ to be piecewise constant over each control step, $\Delta t_c^*$ (taken to be five time steps of the flow solver $\Delta t_c^* = 5\Delta t^*$), and confined to a range $\ddot{\alpha}c^2/U^2_\infty \in [-10,10]$. This range is not based on a specific actuator but it serves to reflect a general physical limitation on actuation in a practical system and to ensure numerical stability during the learning process. Measurements from the system comprise the pressure jump coefficients
\begin{equation}
    C^{j}_{\Delta p} = \frac{P^{j}_l - P^{j}_u}{\frac{1}{2} \rho U_{\infty}^2}, \quad j=1,2,\ldots,10,
    \end{equation}
obtained from $10$ pressure sensors symmetrically distributed on the surface of the flat plate---five placed uniformly between the mid-chord and the edge on each half of the plate---where $P^{j}_l$ and $P^{j}_u$ are the surface pressures on the lower and upper sides, respectively; the lift coefficient
\begin{equation}
    C_L = \frac{F_y}{\frac{1}{2}\rho U_\infty^2 c},
\end{equation}
where $F_y$ is the $y$ component of the force exerted on the flat plate by the surrounding fluid; and the pitching angle $\alpha$, defined as the angle between the chord axis and $x$ axis and defined to be positive in the counter-clockwise direction. Before gusts are introduced, the flat plate has an initial pitching angle $\alpha = 0$ and an initial angular velocity $\dot{\alpha} = 0$.

The flow environment is simulated computationally with a viscous incompressible flow solver based on the immersed boundary projection method \cite{eldredge2022}. The flow solver employs a vorticity-streamfunction formulation and utilizes a lattice Green’s function to solve the streamfunction Poisson equation, allowing for a more compact domain compared to traditional computational aerodynamics studies. From here on, we only report the time in convective time units, $t^{*} = tU_{\infty}/c$, and the spatial coordinates in chord lengths, $x^{*}=x/c$ and $y^{*}=y/c$.  The computational domain spans $[-1.395, 2.175]$ in the $x^*$ direction and $[-0.885, 0.885]$ in the $y^*$ direction and employs free-space conditions on all boundaries. In this computational domain setup, the vortex gust and the resulting airfoil response are ensured to remain between the upper and lower boundaries and freely convect through the downstream boundary. We employ a uniform Cartesian grid with grid size $\Delta x^{*}=0.015$ and a uniform time step size $\Delta t^{*}=0.006$. Each simulation case lasts for $3$ convective time units, but is truncated early if $|\alpha|>\pi/4$ to avoid unrealistic configurations.

\subsection{Reinforcement learning framework}
\label{subsection:rl}
\begin{figure}[htbp]
    \centering
    \includegraphics[width=1.0\linewidth]{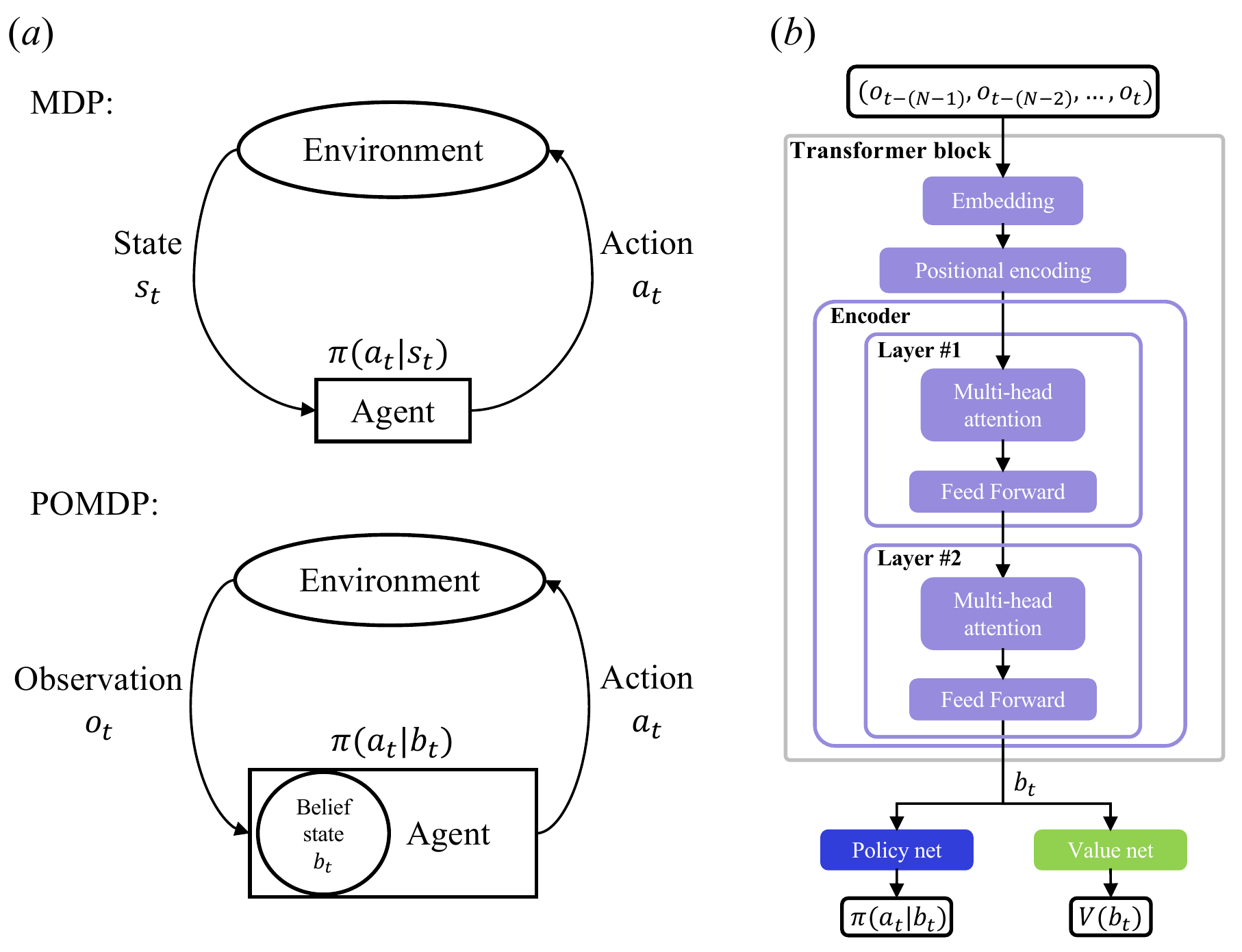}
    \caption{(\textit{a}) Comparison between MDP and POMDP. (\textit{b}) The schematic figure of transformer and the subsequent policy and value networks.}
    \label{fig:pomdp_transformer}
\end{figure}

In this section, we describe how the flow control problem is formulated into a reinforcement learning (RL) framework. As Fig. \ref{fig:pomdp_transformer} shows,
the basic RL framework, consisting of the environment and agent, is expressed as a Markov decision process (MDP). A MDP models the decision-making process by a tuple $\{ \boldsymbol{S}, \boldsymbol{A}, \mathcal{T}, \mathcal{R}, \gamma \}$ where $\boldsymbol{S}$ is the state space, $\boldsymbol{A}$ is the action space, $\mathcal{T}$ is the transition probability function, $\mathcal{R}$ is the reward function, and $\gamma$ is the discounted factor: the agent receives a full state $s \in \boldsymbol{S}$ and selects an action $a \in \boldsymbol{A}$, the state $s$ is transitioned to a new state $s'$ according to the transition probability function $\mathcal{T}(s'|s,a)$, and a reward is also generated from $r=\mathcal{R}(s,a)$. However, the full state $s$ is not accessible in many real world systems, particularly in fluid dynamics, where this state includes the entire flow field. Instead, we only have partial information about the full state $s$ from available sensors, which leads to its formulation as a partial observable Markov decision process (POMDP) \cite{ASTROM1965,KAELBLING1998}. A POMDP extends the MDP tuple to $\{ \boldsymbol{S}, \boldsymbol{A}, \boldsymbol{O}, \mathcal{T}, \mathcal{Z}, \mathcal{R}, \gamma \}$ where, in addition to the MDP components, $\boldsymbol{O}$ is the observation space and $\mathcal{Z} = \mathcal{Z}(o'|s,a)$ is the observation function, representing the probability of receiving observation $o'$ given the state $s$ and action $a$. Therefore, instead of making a decision based on full state $s$ as in a MDP, the agent in a POMDP makes a decision based on a belief state $b(s)$, which represents a probability distribution over all possible states given the agent's observation history. The agent learns a policy $\pi = \pi(a|b(s))$---that is, a policy conditioned on the distribution of possible states---that maximizes the discounted return
\begin{equation}
    J(\pi) = \mathbb{E}\left[ \sum_{t=0}^{\infty}\gamma^t r_t ~|~ \pi \right].
\end{equation}

As we discussed above, the agent relies on a belief state $b(s)$ to make a decision in a POMDP, which requires past observed information. Some RL algorithms propagate this belief state forward in time by folding in observations, and then condition their actions based on the most probable state. This approach is challenging in a fluid dynamics environment, where the state is high dimensional, though there has been recent success in sequentially estimating lower-dimensional representations of the flow state \cite{mousavi2025}. Alternatively, in fluid dynamics, some studies have directly employed the past observation history as a reduced representation of the state to approximate a POMDP as a MDP \cite{Beckers2024,Xia2024,Weissenbacher2025}. Francois-Lavet et al. \cite{francoislavet2019} have shown that the agent can learn to encode the relevant past observations into a latent representation that serves as a surrogate of the probable state and make a decision based on this latent representation. Therefore, sequence models, such as recurrent neural networks (RNNs) \cite{rumelhart1986} and transformers \cite{Vaswani2017}, play an important role since they can extract meaningful information from a time-dependent history sequence. One form of RNN, the long-short term memory (LSTM) \cite{Hochreiter1997}, has been successfully applied to many complex fluid systems to model time-dependent behavior \cite{Liu2025,Hou2019,Hasegawa2020}. The mechanism of the LSTM is to update a hidden state at each time step based on the previous hidden state information and the new observation. This approach has two limitations: it is difficult to retain the information from distinct past observations when the sequence is very long, and the recursive process makes it impossible to process the sequence in parallel.

The limitations of the RNN motivate the recent popularity of the transformer \cite{Vaswani2017}, which leverages the self-attention mechanism to model a time sequence simultaneously. Specifically, the transformer first embeds the observation sequence into a higher dimensional space of tokens using a linear projection and a positional encoding (to ensure that temporal information is included):
\begin{equation}
    f\left(o_{t-(N-1)},o_{t-(N-2)}, \dots, o_t\right) = \left(x_{t-(N-1)},x_{t-(N-2)}, \dots, x_t\right),
\end{equation}
where $o_t\in\mathbb{R}^{\dimobs}$ is the observation at time step $t$, $x_t\in\mathbb{R}^{\dimstate}$ is the embedded token at the corresponding time step, and $N$ is the observation window size, i.e., the observation sequence length. After embedding the observation sequence, the transformer applies the multi-head self-attention mechanism, wherein each embedded token dynamically attends to all $N$ tokens in the sequence by computing the query $\boldsymbol{Q}$, the key $\boldsymbol{K}$, and the value $\boldsymbol{V}$ for each attention head. Each attention head $m = 1, \ldots, \numhead$ uses separate learnable matrices $\boldsymbol{W_Q}^{(m)}, \boldsymbol{W_K}^{(m)}, \boldsymbol{W_V}^{(m)} \in \mathbb{R}^{\dimstate\times \dimhead}$, where $\dimhead=\dimstate/\numhead$, to compute
\begin{equation}
    \boldsymbol{Q}^{(m)} = \boldsymbol{X} \boldsymbol{W_Q}^{(m)}, \qquad \boldsymbol{K}^{(m)} = \boldsymbol{X} \boldsymbol{W_K}^{(m)}, \qquad \boldsymbol{V}^{(m)} = \boldsymbol{X} \boldsymbol{W_V}^{(m)},
\end{equation}
where $\boldsymbol{X} = \left[x_{t-(N-1)} \,\, x_{t-(N-2)} \cdots x_t\right]^{T} \in \mathbb{R}^{N\times \dimstate}$ is a matrix whose rows comprise the $N$ embedded tokens of the input sequence. The $i$th row of $\boldsymbol{Q}^{(m)}$, the query vector $Q_i^{(m)} \in \mathbb{R}^{\dimhead}$, represents what the $i$th token is looking for from the other tokens. The $i$th row of $\boldsymbol{K}^{(m)}$, the key vector $K_i^{(m)}  \in \mathbb{R}^{\dimhead}$, represents the features of the $i$th token that provide for other tokens to attend to. The $i$th row of $\boldsymbol{V}^{(m)}$, the value vector $V_i^{(m)}  \in \mathbb{R}^{\dimhead}$, is the actual information carried by the $i$th token. For each attention head, the scalar-valued attention score $\alpha_{ij}^{(m)}$ between the $i$th and $j$th tokens is computed by
\begin{equation}
    \alpha_{ij}^{(m)} = \frac{Q_i^{(m)}\cdot K_j^{(m)}}{\sqrt{\dimhead}},
\end{equation}
representing how much attention the $i$th token should pay attention to the $j$th token. This attention is used to assemble the output $h_i^{(m)} \in \mathbb{R}^{d_{\text{k}}}$ as the weighted sum of the value vectors $V_j^{(m)}$,
\begin{equation}
    h_i^{(m)} = \sum_{j=1}^{N}\text{softmax}(\alpha_{ij}^{(m)})V_j^{(m)}.
\end{equation}
The final output $h_i \in \mathbb{R}^{\dimstate}$ is obtained by concatenating the $h_i^{(m)}$ from each head into a single vector:
\begin{equation}
    h_i = \text{Concat} \left( h_i^{(1)}, \dots, h_i^{(\numhead)} \right),
\end{equation}
and represents the contextualized representation of the $i$th token. This output of the self-attention layer is fed into the feedforward network layer, which introduces nonlinearity to refine the token representation. The combination of the self-attention layer and the feedforward layer is an encoder layer. In the current work, our transformer employs two encoder layers with two attention heads in each layer. After the observation sequence $\left(o_{t-(N-1)},o_{t-(N-2)}, \dots, o_t\right)$ is processed by the two-layer transformer structure, we obtain a sequence of contextualized representations $\left( \tilde{h}_{t-(N-1)}, \tilde{h}_{t-(N-2)}, \dots, \tilde{h}_t \right)$, where $\tilde{h}_t \in \mathbb{R}^{\dimstate}$ incorporates information from the entire observation window. In the current work, we extract the final element $\tilde{h}_t$ from this sequence and treat it as the latent representation of the probable state. This latent representation is then passed to the policy and value networks, which provide the probability distribution over the action space (for action selection) and the approximated expected return on that state (for state-value estimation), respectively. Fig. \ref{fig:pomdp_transformer}(b) summarizes this framework. For clarity, $\dimobs=12$, $\dimstate=32$, $\dimhead=16$, and $\numhead=2$ in this study. \revision{We also note that the adopted transformer structure in the present study is empirically demonstrated as a practical choice but may not necessarily be the only choice, and a comprehensive architecture benchmark (e.g. transformer vs. LSTM vs. MLP) will be an interesting direction for future work.}

We use the proximal policy optimization (PPO) algorithm \cite{schulman2017} to approximate the optimal policy. Both the policy and the value networks consist of two hidden layers with $32$ hidden units each, where each hidden layer employs a Rectified Linear Unit activation function. The loss function that training seeks to minimize comprises three components. PPO, as a gradient-based model-free algorithm following the actor-critic framework, introduces a clipped surrogate objective to ensure updating within a safe region, constituting the first loss component
\begin{equation}
    L^{\text{policy}}_{\theta} = \mathbb{E}_t \left[ \text{min}(\rho_t A_t, \text{clip}(1,1-\epsilon,1+\epsilon)A_t) \right],
\end{equation}
where $\rho_t = \pi_{\theta}(a_t|s_t)/\pi_{\theta_{\text{old}}}(a_t|s_t)$ is the ratio of updated and old policies; $A_t$ is the advantage function, which helps the agent evaluate how good the action $a_t$ is at state $s_t$ compared to the expected value of this state $V(s_t)$; and $\epsilon$ is the clip hyperparameter that controls the size of the safe region. The advantage function is estimated from the generalized advantage estimation (GAE) \cite{schulman2018}, defined as
\begin{equation}
    A^{\text{GAE}}_t = \sum_{l=0}^{\infty}(\gamma \lambda)^l \delta_{t+l},
\end{equation}
where $\delta_t = r_t + \gamma (1-d_t)V_{\phi}(s_{t+1}) -V_{\phi}(s_t)$ is the temporal difference residual, and $\lambda$ is a hyperparameter that controls the trade-off between the variance and bias of the estimator. The variable \( d_t \in \{0,1\} \) is a Boolean terminal signal indicating whether the next state \( s_{t+1} \) is terminal (i.e., \( d_t = 1 \)) or not (i.e., \( d_t = 0 \)). Furthermore, the value loss function is defined as
\begin{equation}
    L^{\text{value}}_{\phi} = \mathbb{E}_t \left[\left(r_t + \gamma V_{\phi}(s_{t+1}) - V_{\phi}(s_t) \right)^2 \right],
\end{equation}
where $\gamma$ is the discounted factor. For encouraging exploration, an entropy loss is incorporated
\begin{equation}
    L^{\text{entropy}} = \mathbb{E}_t \left[ - \sum_{a_t} \pi(a_t|s_t)\log\pi(a_t|s_t) \right].
\end{equation}
Finally, we formulate the total loss as
\begin{equation}\label{eq:loss}
    L(\theta,\phi) = L^{\text{policy}}_{\theta} + C_V L^{\text{value}}_{\phi} - C_E L^{\text{entropy}},
\end{equation}
where $C_V$ is the value coefficient and $C_E$ is the entropy coefficient. 

RL training requires a large number of interactions between the agent and the environment (in this paper, CFD) if it starts from scratch, and thus the application of RL may be impeded in expensive environments (e.g., flow control around a three-dimensional wing shape). Therefore, in this paper we will seek to accelerate training by using a pretraining technique to initialize a structured policy from expert data, i.e., actions obtained from a prior control strategy that performs reasonably well. In the present work, this prior strategy is P control, and the pretrained policy is derived by sampling actions on trajectories that follow the proportional control strategy. Specifically, we find an initial policy that minimizes a loss function as follows:
\begin{equation}
    L_{\theta}^{\text{pretraining}} = \mathbb{E}_{\mathcal{D}} \left[ a_{\text{expert}} - \hat{a} \right],
\end{equation}
where $a_{\text{expert}}$ is the expert action, i.e., the target, collected from the sampled expert trajectories $\mathcal{D}$, and $\hat{a}$ is the output from the transformer policy network.

\begin{figure}[htbp]
    \centering
    \includegraphics[width=1.0\linewidth]{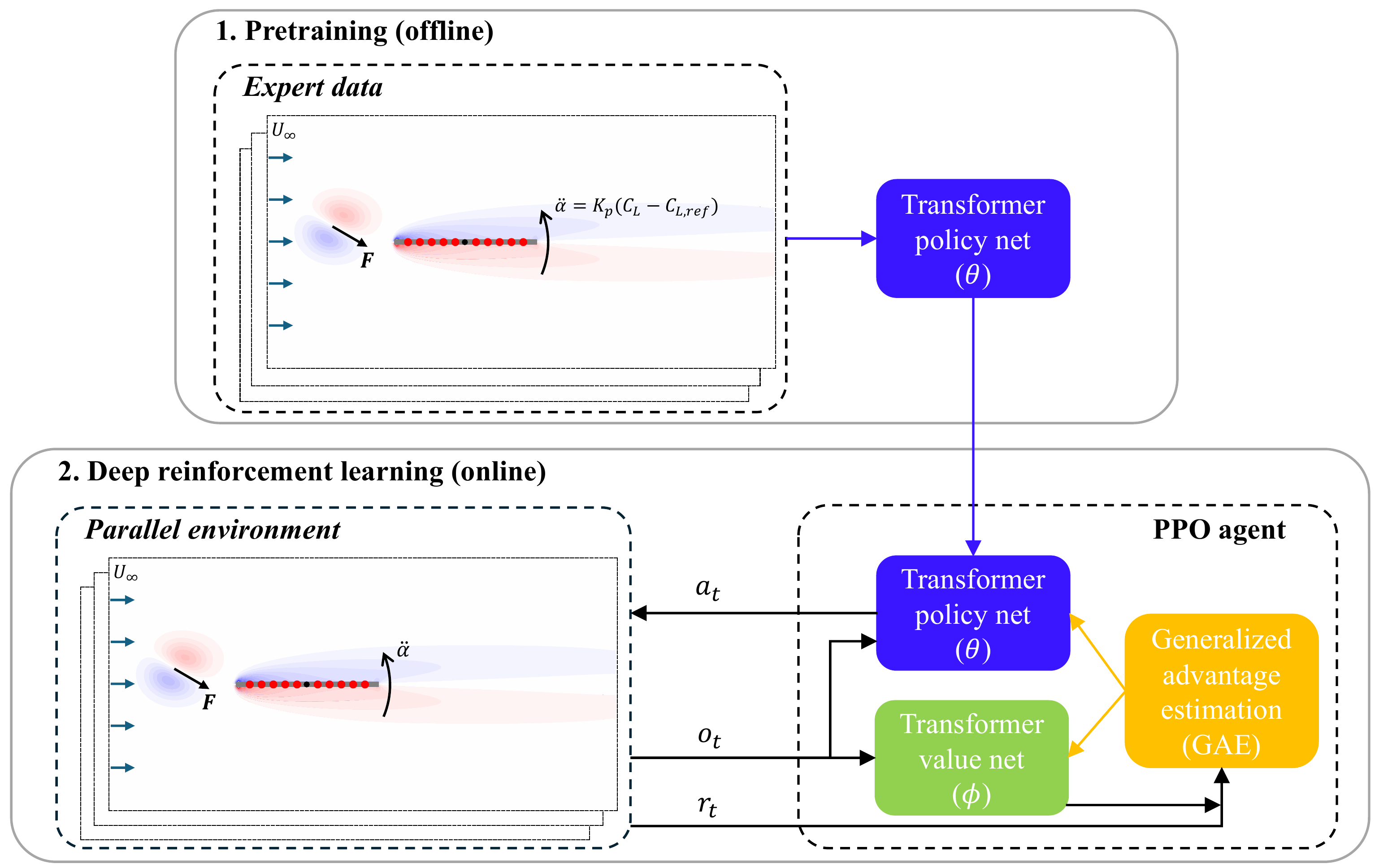}
    \caption{Overview of the transformer-based RL framework}
    \label{fig:workflow}
\end{figure}

We summarize the overall transformer-based RL framework in Fig. \ref{fig:workflow}.
From the initial policy network, which may be random or initialized by pretraining, we move on to the RL training. A parallel environment is employed to enhance the data efficiency. As described above, the action $a_t$ in this paper is defined as the angular acceleration $\ddot{\alpha}$ about a fixed pivot point and the observation $o_t$ is defined as a vector consisting of $10$ pressure sensor measurements over the surface of the flat plate with the locations indicated by the red dot in Fig. \ref{fig:workflow}, the lift coefficient, and the pitching angle:
\begin{equation}
    o_t = \left[C_{\Delta p}^1\cdots C_{\Delta p}^{10}\,\,C_L\,\,\alpha \right]_t.
\end{equation}
From previous work \cite{mousavi2025}, it has been shown that surface pressure can be informed effectively by the change of flow field, and thus we incorporate surface pressure into the observation function. An observation window with a size of $N=20$ is employed throughout this study. \revision{A comparison between the RL agents trained with observation window sizes $N=20$ and $N=40$, as shown in appendix \ref{appendix}, suggests that $N=20$ is an efficient and sufficient choice in the present study.} Since we aim to find an optimal policy that enables the agent to track a reference lift profile, we design the reward function as follows:
\begin{equation}\label{eq:reward}
r_t = 1 - \left( \frac{C_{L_t}-C_{L_{\text{ref}}}}{C_{L_{\text{scale}}}} \right)^2 
      - \left| \frac{\ddot{\alpha}_{t}-\ddot{\alpha}_{t-1}}{\ddot{\alpha}_{\text{scale}}} \right| 
      - \begin{cases}
          100, & \text{if truncated}, \\
          0, & \text{otherwise}.
        \end{cases}
\end{equation}
The first term serves as a baseline reward at each time step, ensuring that the maximum reward under an ideal control strategy is $1$, facilitating stable learning. The second term in \eqref{eq:reward} penalizes the deviation of the lift coefficient $C_L$ from the reference lift $C_{L_{\text{ref}}}$, with the denominator $C_{L_{\text{scale}}}$ included to scale this term to an appropriate weight in the overall reward; a quadratic form is chosen to amplify large deviation compared to a linear penalty, and this choice is confirmed from empirical observations that it significantly improves the learning performance, likely due to a smoother gradient behavior during the policy optimization. The third term penalizes temporal changes of the control input, encouraging a smoother control process and avoiding an unrealistically aggressive actuation; the scale factor $\ddot{\alpha}_{\text{scale}}$ modifies its relative importance in the total reward. The final term imposes a large penalty on episodes that truncate prematurely due to exceeding the feasible range of pitching angle, i.e. $|\alpha|>\pi/4$. The reference lift in this study is set to zero, $C_{L_{\text{ref}}}=0$, and based on experience, we set $C_{L_{\text{scale}}}=0.3$ and $\ddot{\alpha}_{\text{scale}}=10$ throughout the study.

\begin{figure}[htbp]
    \centering
    \includegraphics[width=1.0\linewidth]{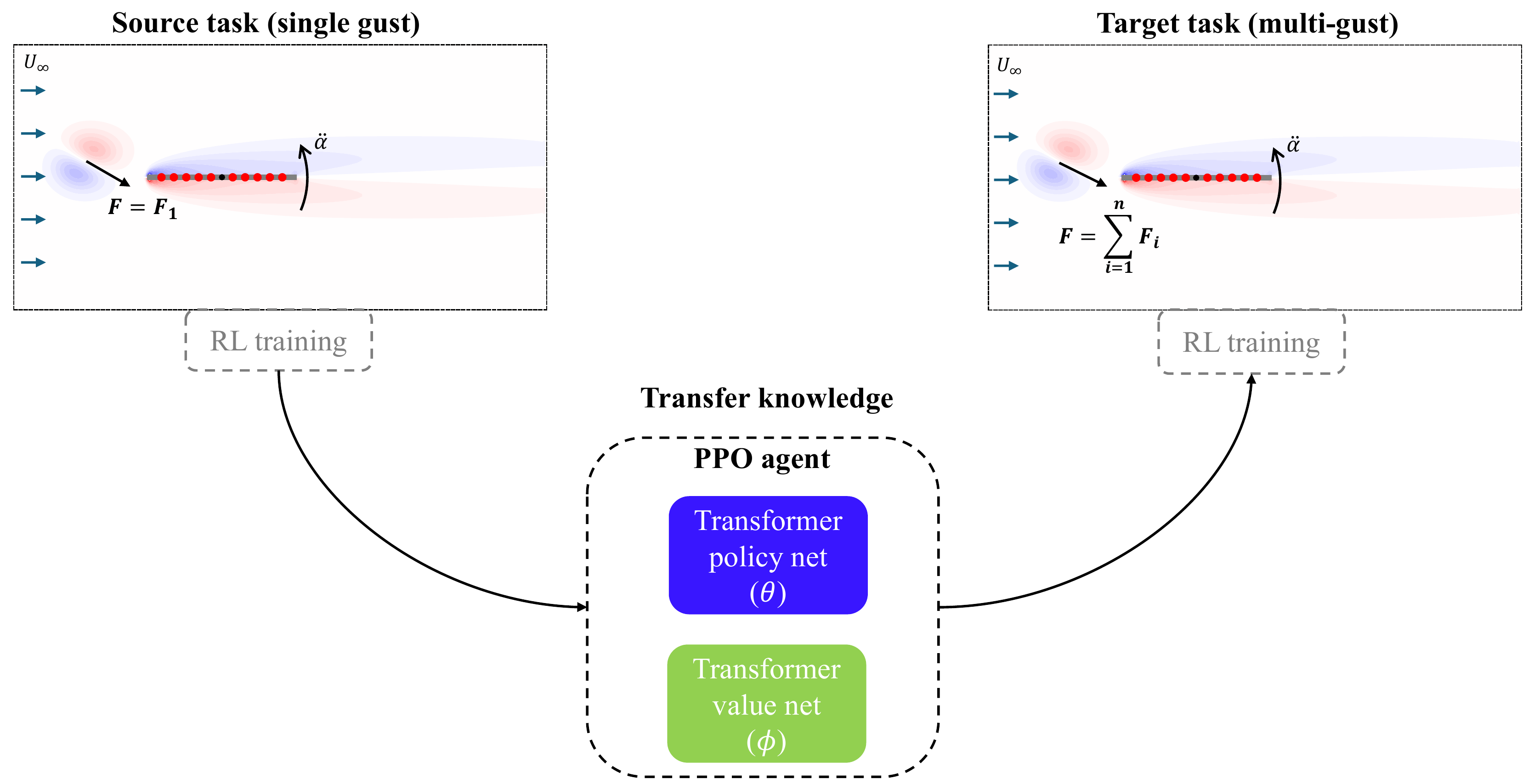}
    \caption{Overview of the transfer learning.}
    \label{fig:transfer_learning}
\end{figure}

Before we complete this description of the methodology, we described one more technique that we use in this paper to accelerate the training. Finding a good expert policy from which to initiate training is often nontrivial. Therefore, we leverage task-level transfer learning \cite{bozinovski2020} as another strategy when the expert data for the pretraining is not available. Transfer learning has been successfully applied to fluid dynamics problems recently \cite{Wang20222,YAN2025}. As depicted in Fig. \ref{fig:transfer_learning}, the RL agent is first trained in a single-gust environment (the source task), and once it acquires sufficient knowledge, the learned policy and value networks are transferred to a multi-gust environment (the target task), thereby accelerating learning and reducing training cost. This approach facilitates the application of RL to environments in which training is even more expensive: by first training a RL agent in a relatively easy or simplified environment, we obtain a well-informed initialization that serves as a warm start in the subsequent RL training in a more complex environment, and thus the overall training cost may be reduced.

\section{Results and discussion}
\label{sec:resultsanddiscussion}

In this section, we demonstrate the training, performance, and generalizability of the proposed transformer-based RL framework on the lift regulation problem. The first part of the study, in Section \ref{subsec:benchmarking} and Section \ref{subsec:extending}, utilizes the mid-chord pitching configuration. In Section \ref{subsec:benchmarking} we benchmark the performance of P control and RL control in a baseline environment, comprising a single gust. We then investigate the generalizability of the RL design approach in Section \ref{subsec:extending} by extending to an environment that includes multiple gusts. Then, we focus attention on the quarter-chord pitching configuration in Section \ref{subsec:revisiting} and Section \ref{subsec:transferring}. We first study the effect of this change of pivot point location in the single-gust environment in Section \ref{subsec:revisiting}, devoting particular attention to the physics of the two different configurations. Finally, we investigate the extension of the RL control strategy to a multi-gust environment by transfer learning in Section\ref{subsec:transferring}. For reproducibility, the essential hyper-parameters for all configurations and gust environments are summarized in Table \ref{tab:parameter_learning}.

\begin{table}[b]
\caption{Summary of the essential hyper-parameters for all different configurations.}
\label{tab:parameter_learning}
\begin{ruledtabular}
\begin{tabular}{ccccccccc}
    & \makecell{learning \\ rate} & \makecell{clipping \\ ratio} & \makecell{value \\ coefficient} & \makecell{entropy \\ coefficient} & \makecell{number of \\ parallel envs.} & \makecell{inner \\ epochs} & \makecell{discount \\ factor} & \makecell{GAE \\ lambda} \\
    \hline
      \makecell{scratch \\ (single)} & $0.0001$   & $0.2$ & $0.5$   & $0.005$ & $24$ & $24$ & $0.99$ & $0.95$ \\
      \makecell{pretrain \\ (single)} & $0.0001$   & $0.1$ & $0.5$   & $0.005$ & $24$ & $8$ & $0.99$ & $0.95$ \\
      \makecell{scratch \\ (multiple)} & $0.0001$   & $0.2$ & $0.5$   & $0.005$ & $72$ & $72$ & $0.99$ & $0.95$ \\
      \makecell{pretrain \\ (multiple)} & $0.0001$   & $0.1$ & $0.5$   & $0.005$ & $72$ & $8$ & $0.99$ & $0.95$ \\
\end{tabular}
\end{ruledtabular}
\end{table}

\subsection{Benchmarking control performance: mid-chord pitching in single-gust environment}
\label{subsec:benchmarking}
We begin by investigating a baseline environment, which is characterized by a single upstream gust. This configuration is set up to be a benchmark for comparing the performance of P control against RL control. Specifically, we intend to evaluate how well each approach can mitigate the lift variation induced by the gust, and to establish a performance baseline before we move on to a more complex environment. The P control strategy, also explored in previous work as a baseline for RL \cite{Beckers2024}, is defined as follows:
\begin{equation}
\label{eq:Pcontrol}
    \ddot{\alpha} = K_p (C_L - C_{L_{\text{ref}}}),
\end{equation}
where $K_p$ is the proportional control gain to be determined. Rather than train a RL agent from scratch, which requires a significant number of interactions, we hypothesize that we can leverage a P control strategy with a good $K_p$ value to pretrain the RL agent and continue the subsequent RL training from this warm start in order to make the learning process converge faster. 

\begin{figure}[htbp]
    \centering
    \begin{overpic}[width=1.0\linewidth]{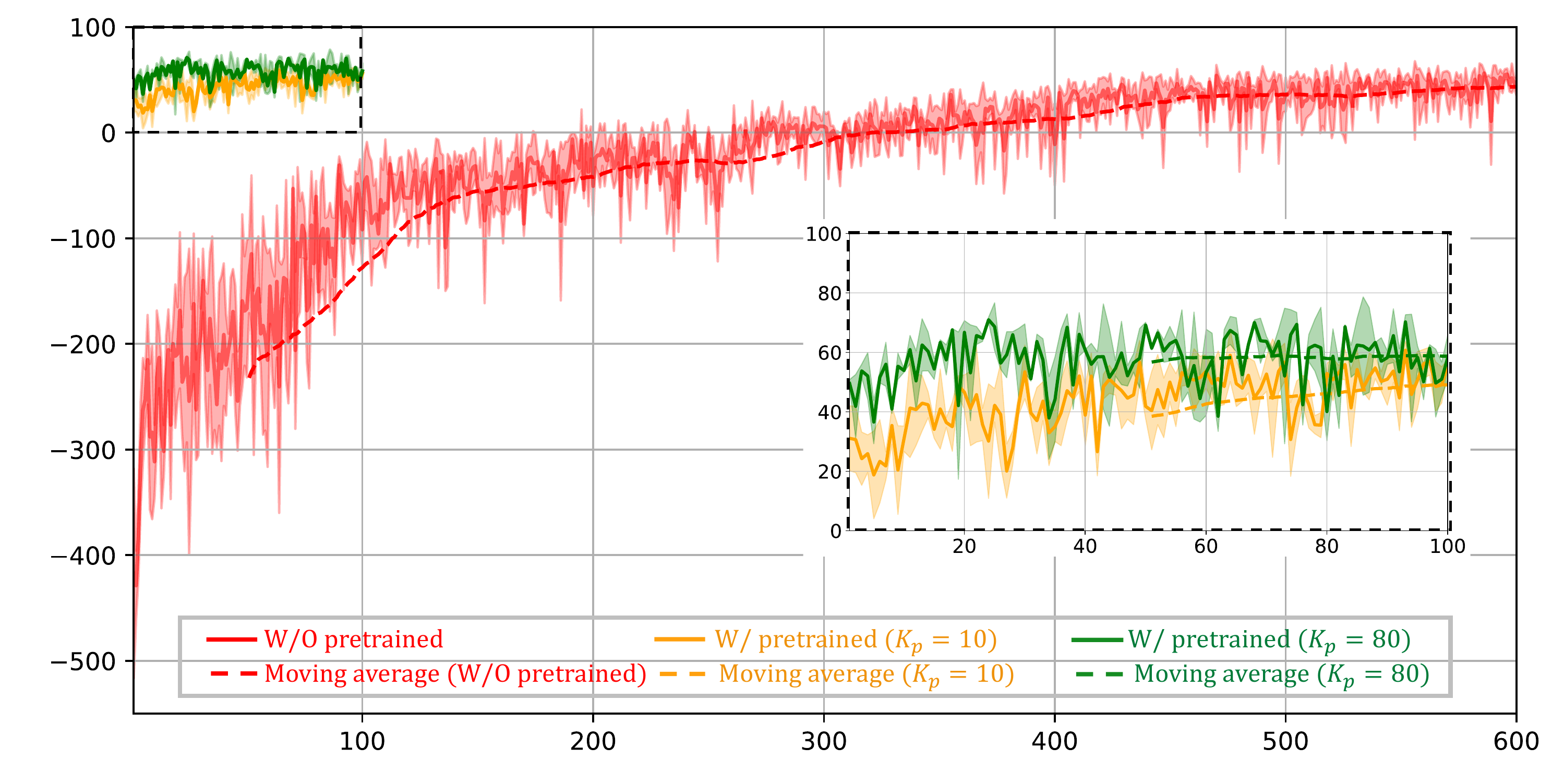}
    \put(0,17){\rotatebox{90}{Episode Rewards}}
    \put(49,-0.5){{Episodes}}
    \put(49.5,18){\rotatebox{90}{\small Episode Rewards}}
    \put(70,12){{\small Episodes}}
    \end{overpic}
    \caption{Learning envelopes for mid-chord pitching RL agents in single-gust environment. The solid lines and the shadow regions are the mean and standard deviation across three different runs. The dashed lines show the moving average with an averaging window of $50$ episodes. The inset plot is a zoomed-in view for the learning envelopes of RL agents starting from pretrained models. At each episode index, the episode reward is averaged across all parallel environments per run.}
    \label{fig:learning_mid_single}
\end{figure}
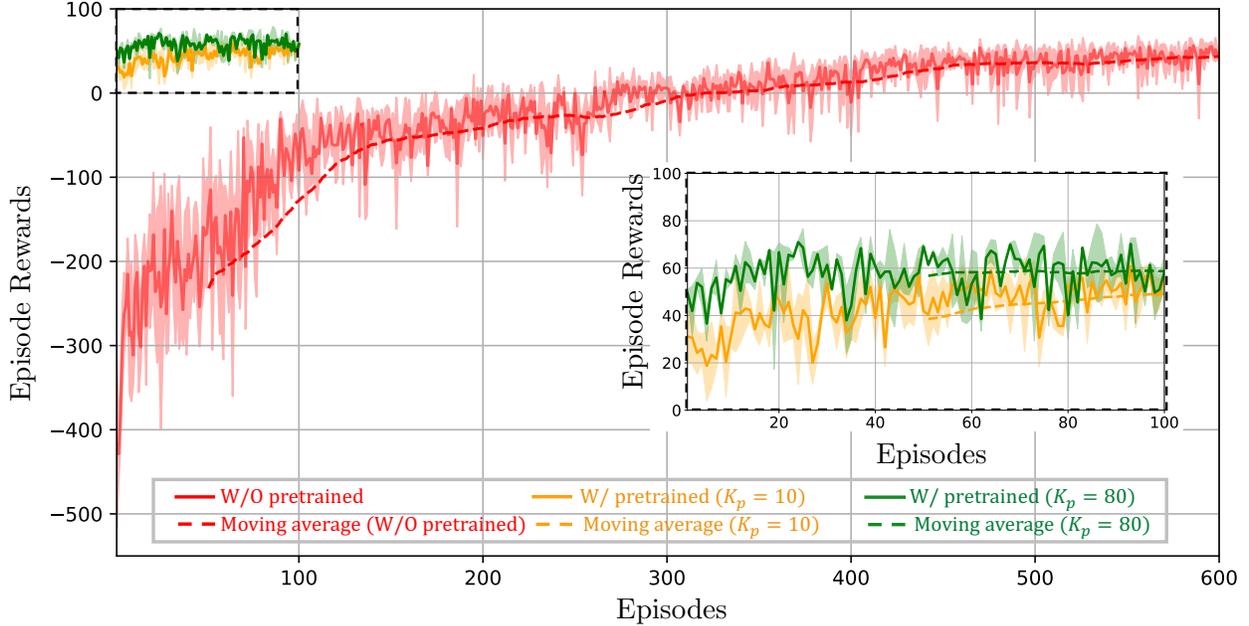

We show the learning envelope for the RL agents starting from scratch and those pretrained with P control in Fig. \ref{fig:learning_mid_single}. We report the total reward $\sum r_t$ in each episode as the criterion to evaluate the performance of the RL agent evolves with the number of training episodes. We can first observe that the red learning envelope, which corresponds to learning from scratch, starts with an episode reward of approximately $-430$, reflecting the poor performance of an uninformed random policy. An episode total reward of $0$ is not reached until around the $300\text{th}$ episode, there is substantial variance throughout the learning process, and the learning envelope is still not fully converged until the $600\text{th}$ episode. This behavior reflects the high cost of exploration typically required for the agent to discover a viable control strategy. In contrast, when we start with a well-informed pretrained policy, as in both the orange and green learning envelopes, we have a much higher initial episode total reward and the training curve exhibits significantly less variance. This highlights the benefit of pretraining in enabling the agent to avoid extensive and inefficient exploration. Among the two pretrained strategies, a pretrained policy obtained from a P control strategy with $K_p=80$ (in green) has a higher initial total reward than the one that is pretrained with $K_p=10$ (in orange), and thus, the former is already converged at the $100\text{th}$ episode, whereas the latter is still converging, though likely to finish in modestly more episodes. Therefore, these results show that strong expert data for pretraining helps the RL agent converge faster.

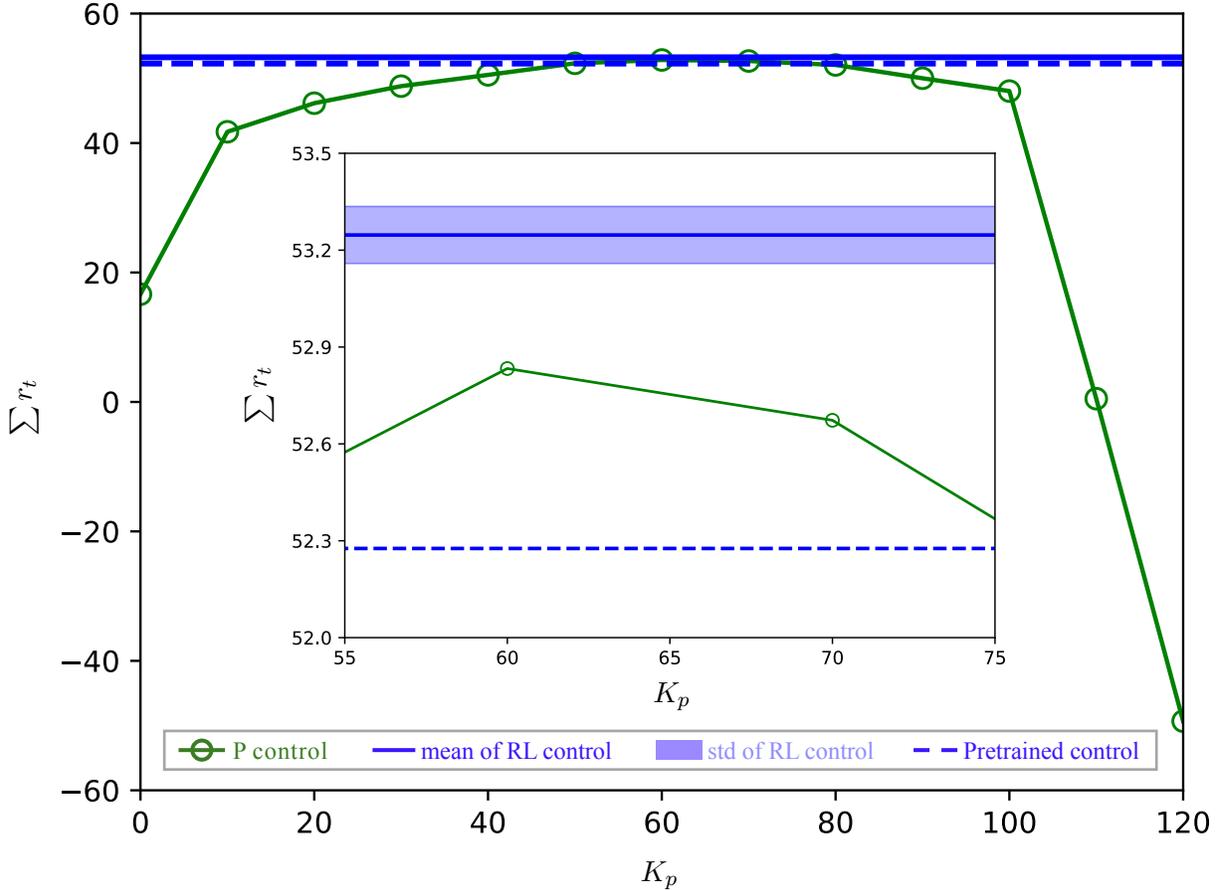
\begin{figure}[htbp]
    \centering
    \begin{overpic}[width=1.0\linewidth]{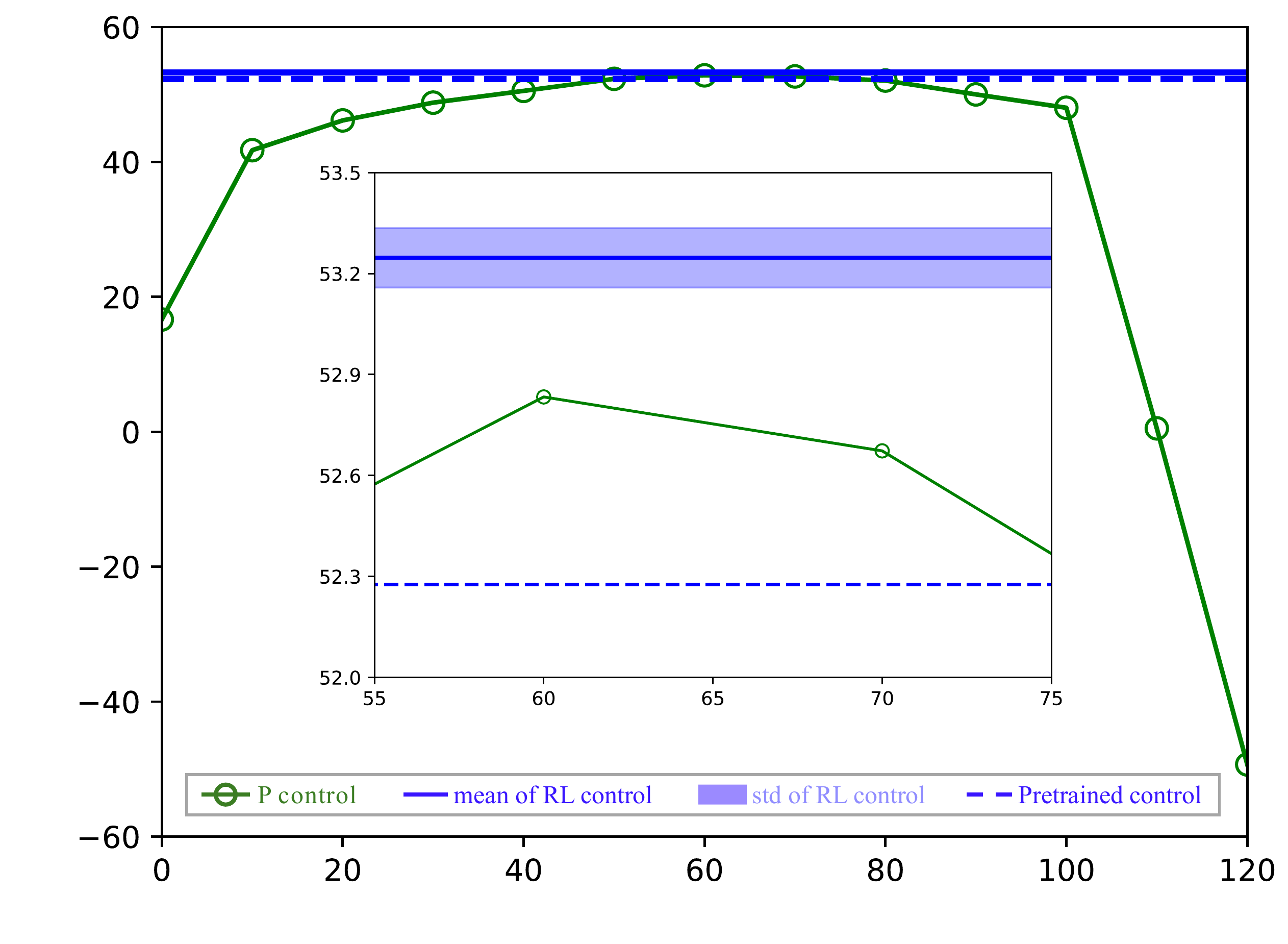}
    \put(2,37){\rotatebox{90}{$\sum r_t$}}
    \put(53,1){{$K_p$}}
    \put(21,38){\rotatebox{90}{$\sum r_t$}}
    \put(54,15.5){{$K_p$}}
    \end{overpic}
    \caption{Evaluation of proportional control and RL control in single-gust environment. The statistics for each data point is averaged over $15$ random cases. The blue solid line and shadow region represent the mean and standard deviation value for the RL agents in the last $5$ episodes. The dashed line represents the pretrained RL agent. The inset figure is a zoom-in view for $K_p$ from $55$ to $75$.}
    \label{fig:eval_single_mid}
\end{figure}

For further evaluation we focus on the fully-trained RL agent initialized from a pretrained P control policy with $K_p=80$. To assess the policy's robustness, RL agents are picked from the last five episodes. In Fig. \ref{fig:eval_single_mid} we compare these agents with control strategies based only on P control. It should be noted that, although the P control only penalizes the instantaneous lift error ($C_L - C_{L_{\text{ref}}}$), we evaluate its performance using the same reward function employed during the RL training. This reward function is a more comprehensive and practical evaluation metric, incorporating the lift track error but also penalizing aggressive control input and unrealistic configurations, thereby balancing the precision with smoothness and stability. Thus, we acknowledge that there is a slight mismatch in the evaluation basis. Later in this section, we will compare the different control strategies in individual episodes on the basis of lift variation alone.

From Fig. \ref{fig:eval_single_mid}, we can observe that starting from $K_p=0$, i.e. no control, the performance of the P control improves with increasing gain until $K_p=60$, where it shows the optimal performance of all the P control strategies. Beyond that value of $K_p$, the P control starts to deteriorate, particularly after $K_p=100$, indicating that an excessively high gain causes instability and an overreaction to the gust disturbances. We can observe that the envelope of trained RL control strategies (shown as a blue envelope) improve upon the performance of the initial pretrained policy (shown as a blue dashed line) and outperform the best P control across the tested $K_p$ region. However, the performance gap between the mean of RL control and the best P control is only approximately $0.4$, showing that linear control is an effective control strategy for single-gust episodes mitigated by mid-chord pitching and RL does little to improve on it. We expect this performance gap will become larger as the number of gusts increases which would introduce higher nonlinearity, as we will discuss in Section \ref{subsec:extending}.

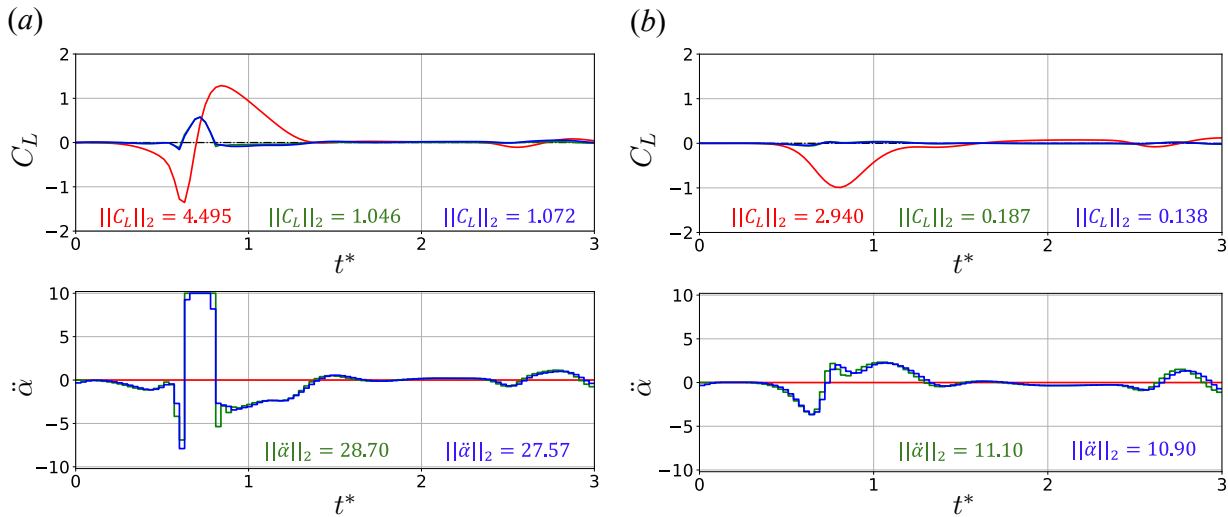
\begin{figure}[htbp]
    \centering
    \begin{overpic}[width=1.0\linewidth]{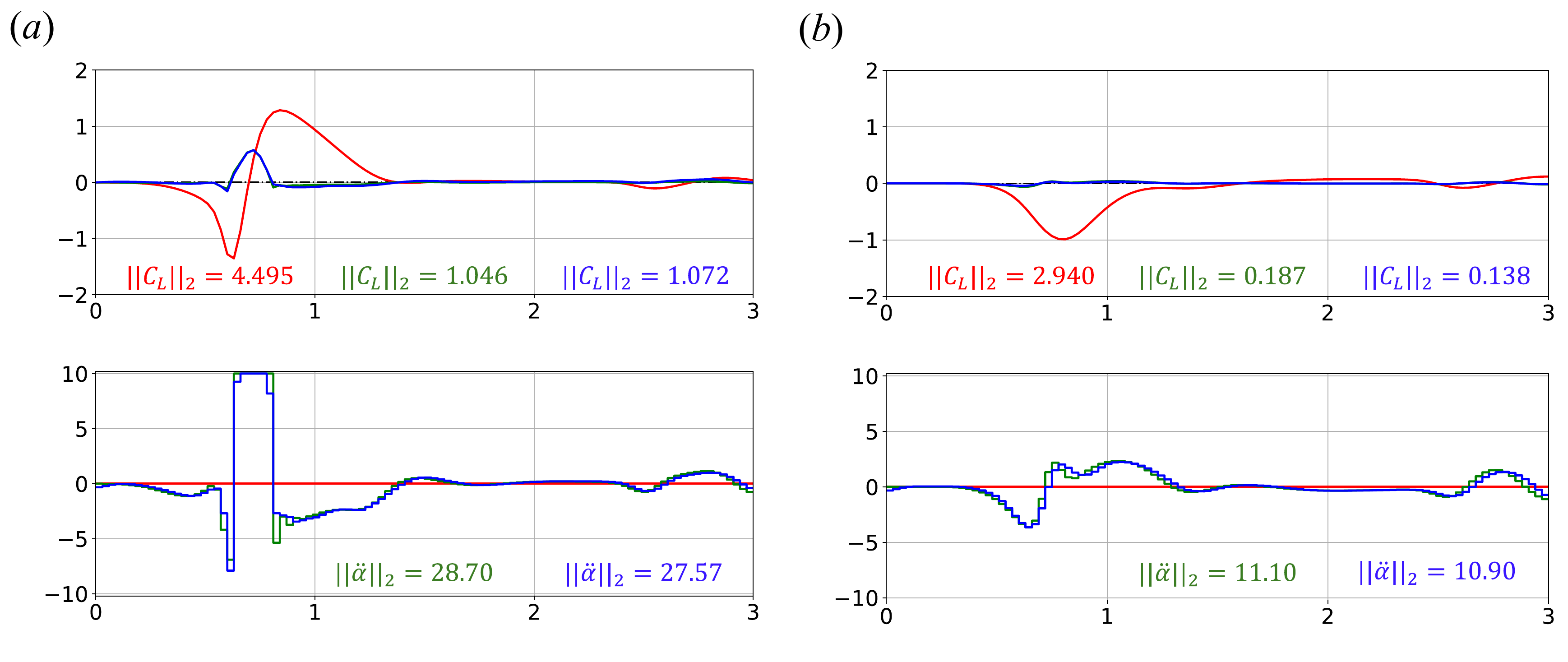}
    \put(1,28.5){\rotatebox{90}{$C_L$}}
    \put(27,20){{$t^*$}}
    \put(51,28.5){\rotatebox{90}{$C_L$}}
    \put(77,20){{$t^*$}}
    \put(1,10){\rotatebox{90}{$\ddot{\alpha}$}}
    \put(27,0.5){{$t^*$}}
    \put(51,10){\rotatebox{90}{$\ddot{\alpha}$}}
    \put(77,0.5){{$t^*$}}
    \end{overpic}
    \caption{The lift histories and angular acceleration histories under no control, P control ($K_p=60$), and RL control for representative single-gust case 1 in (\textit{a}), and representative single-gust case 2 in (\textit{b}). The color convention is \textcolor{red}{\rule[0.5ex]{1.5em}{1pt}: no control}, \textcolor{green!60!black}{\rule[0.5ex]{1.5em}{1pt}: P control},
\textcolor{blue}{\rule[0.5ex]{1.5em}{1pt}: RL control}. The corresponding total rewards are: (\textit{a}) \textcolor{red}{$\sum r_t=-124.54$}, \textcolor{green!60!black}{$\sum r_t=82.42$}, \textcolor{blue}{$\sum r_t=82.65$}; (\textit{b}) \textcolor{red}{$\sum r_t=3.96$}, \textcolor{green!60!black}{$\sum r_t=97.40$}, \textcolor{blue}{$\sum r_t=97.67$}.}
    \label{fig:lift_compare_single_mid}
\end{figure}

To demonstrate the performance of the different control strategies, we show the corresponding lift profile and the action profiles for two representative cases in Fig. \ref{fig:lift_compare_single_mid}. For each representative case, we also report the corresponding $L_2$ norm of the lift coefficient history $||C_L||_2$ (to measure the lift tracking error) and the action history $||\ddot{\alpha}||_2$ (to quantify the total control effort), as well as the episode reward $\sum r_t$. For clarity, the P control here corresponds to $K_p=60$ due to its optimal performance in Fig. \ref{fig:eval_single_mid}. For both representative cases, we observe that RL control shows an almost identical lift tracking error to the best P control, but requires a smaller control effort, and thus the episode reward of the RL control is slightly higher than the best P control. This performance difference is also apparent in the profiles. Specifically, the RL control returns a $C_L$ history that is almost indistinguishable from the best P control, but the $\ddot{\alpha}$ histories behave visibly differently. For example, we can see from the representative case 1 that the RL control learns to smooth out the small temporal fluctuation of $\ddot{\alpha}$ at around $0.75$ convective time. The results of the two representative cases suggest that a well-tuned P control can perform effectively in this relatively simple configuration. In this sense, the RL agents act as a fine-tuned version of the P controller, optimizing the performance in a nuanced manner.

By comparing the two representative cases, we also see that the $C_L$ profile under RL control still shows a small bump in representative case 1, whereas it is not present in representative case 2. This discrepancy can be attributed to the magnitude constraint we impose on the angular acceleration $\ddot{\alpha}$. This is clear from the $\ddot{\alpha}$ history, in which we can see that the time corresponding to the small bump in representative case 1 coincides with the time at which the $\ddot{\alpha}$ reaches the magnitude constraint and also with the interval that the gust induces a huge lift variation on the plate without control. In contrast, in representative case 2, the $\ddot{\alpha}$ history remains well within the bound, and the lift variation without control remains relatively moderate throughout the episode. As a result, the corresponding $C_L$ history under RL control tracks the zero reference lift very accurately. This observation highlights that the actuation limit can restrict the performance of the controller to fully compensate for the strong gust when we implement pitching about the mid-chord. This insight motivates a further investigation about the pivot point location, as we will discuss in Section \ref{subsec:revisiting}.

\revision{In principle, the proportional controller used in this study could be extended to a proportional-integral-derivative (PID) controller, and its gains could be tuned, either automatically via some optimization algorithms or manually, against the same reward function used in our RL setting, but each evaluation of a single set of PID gains requires many simulations across different gust scenarios (varying gust parameters and numbers of gust), and a tuning process of PID gains requires many such evaluations, so a systematic optimization of PID gains for all cases considered in this study entails a substantial additional computational effort. Since the focus of this study is not about an optimal PID controller design but rather to compare a simple but well-performing baseline with our RL controller, we restrict attention to a tuned proportional controller and leave a more exhaustive study about PID (and other linear) baselines for future work.}

\subsection{Extending control capabilities: mid-chord pitching in multi-gust environment}
\label{subsec:extending}
Having shown the effectiveness of the P control and RL control in a relatively simple environment involving a single gust, we now turn to a more realistic and challenging scenario, characterized by successive gusts. In an ideal setting, one would consider an environment with an infinite sequence of gusts to fully capture the complexity of the real-world disturbances. However, such a setup is computationally infeasible for training, so a reasonable choice of a finite number is necessary. Therefore, we choose to first train the RL agent in environments with up to three successive gusts, and then evaluate the control strategy in environments with up to eight gusts as an approximation of an infinite successive gust environment. This setup is intended to address the following key questions: Does the performance gap between the RL control and the P control  widen in the multi-gust environment compared to the gap observed in the single-gust environment, and if so, in what manner? How well do RL control strategies trained in single- or three-gust environments generalize to the more complex eight-gust setting, and does training in a multi-gust environment yield superior performance? By examining these questions, we aim to assess the capability and scalability of the RL control strategies.

\begin{figure}[htbp]
    \centering
    \begin{overpic}[width=1.0\linewidth]{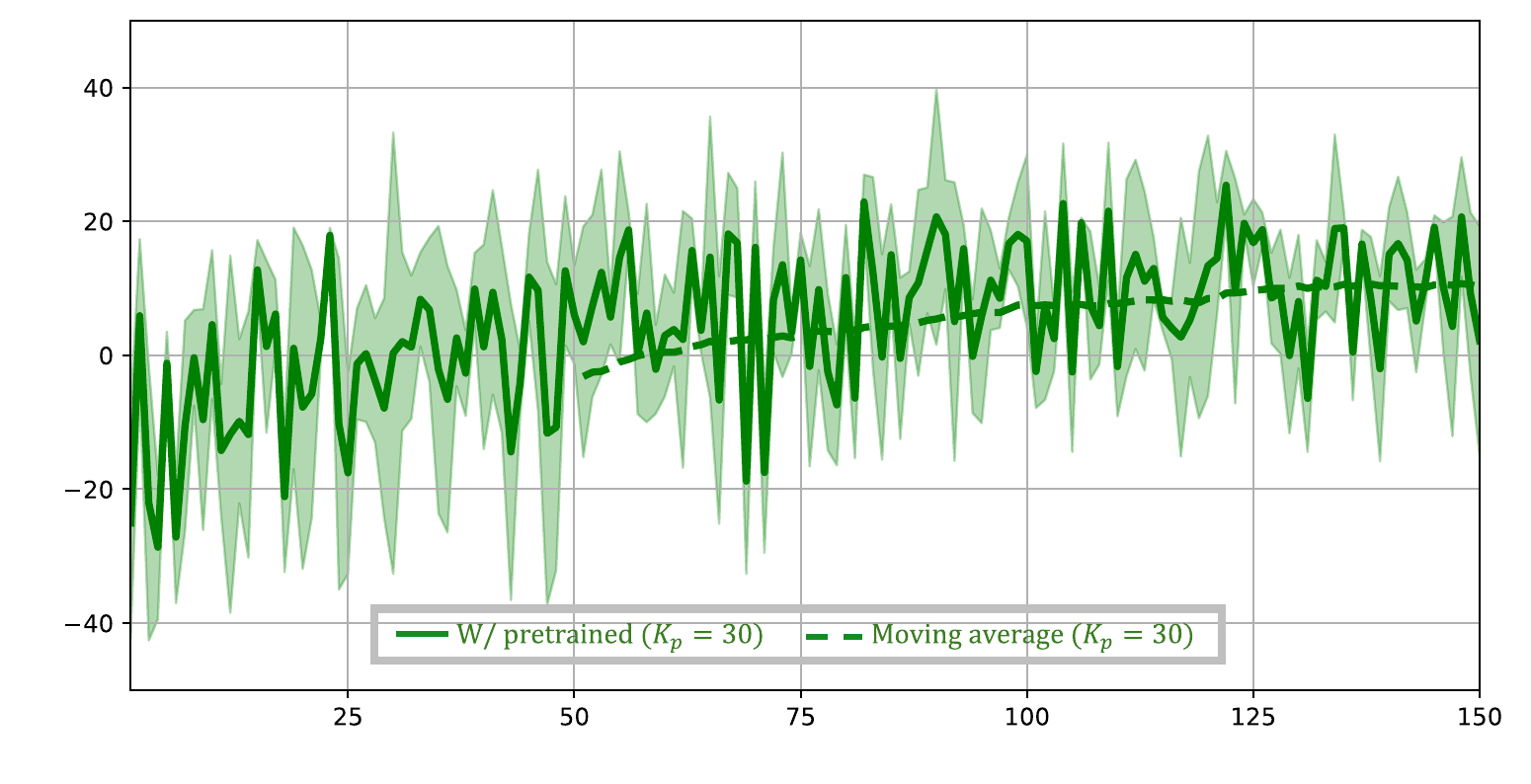}
    \put(1,19){\rotatebox{90}{Episode Rewards}}
    \put(48,1){{Episodes}}
    \end{overpic}
    \caption{Learning envelopes for mid-chord pitching RL agents in three-gust environment. The solid lines and the shadow regions are the mean and std values of the episode sum rewards across three different runs. For the dashed lines, the moving average has a window size of $50$. At each episode index, the episode reward is averaged across all parallel environments per run.}
    \label{fig:learning_mid_three}
\end{figure}

The learning envelopes for the RL agent trained in a three-gust environment is depicted in Fig. \ref{fig:learning_mid_three}. Here, P control with gain $K_p=30$ is selected for pretraining, but we note that other $K_p$ values may also serve effectively as pretraining expert data. We can observe that the learning starts at an initial reward of around $-25$, exhibits a steady improvement, and appears to converge by the $150\text{th}$ episode. By comparing with the learning envelopes in a single-gust environment, the present training curve exhibits a notably larger variance, especially during the early episodes. This variance is reasonable due to the increased complexity and stronger disturbance introduced by three gusts. The converged training curve shows that the RL agent is capable of learning an effective control strategies in a more complex environment, emphasizing the generalizability of the proposed learning framework.

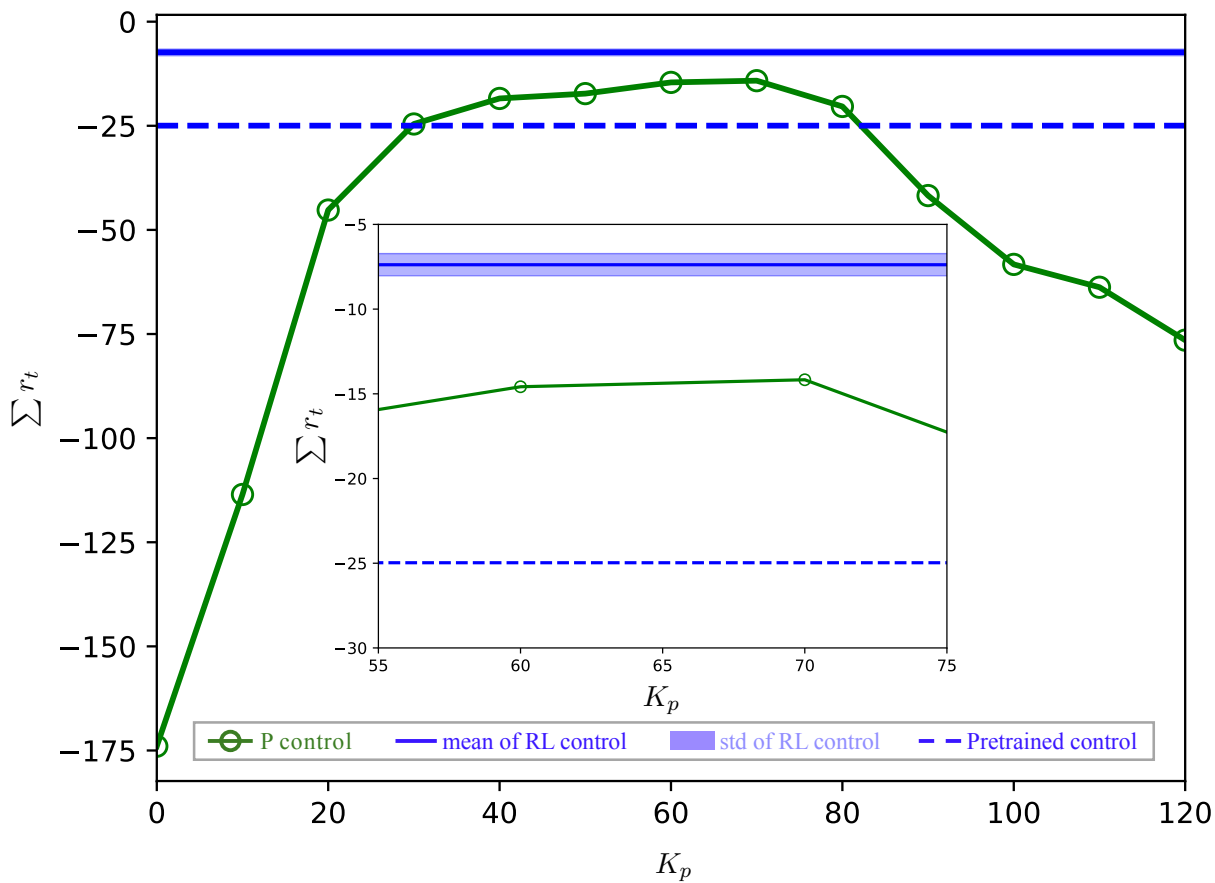
\begin{figure}[htbp]
    \centering
    \begin{overpic}[width=1.0\linewidth]{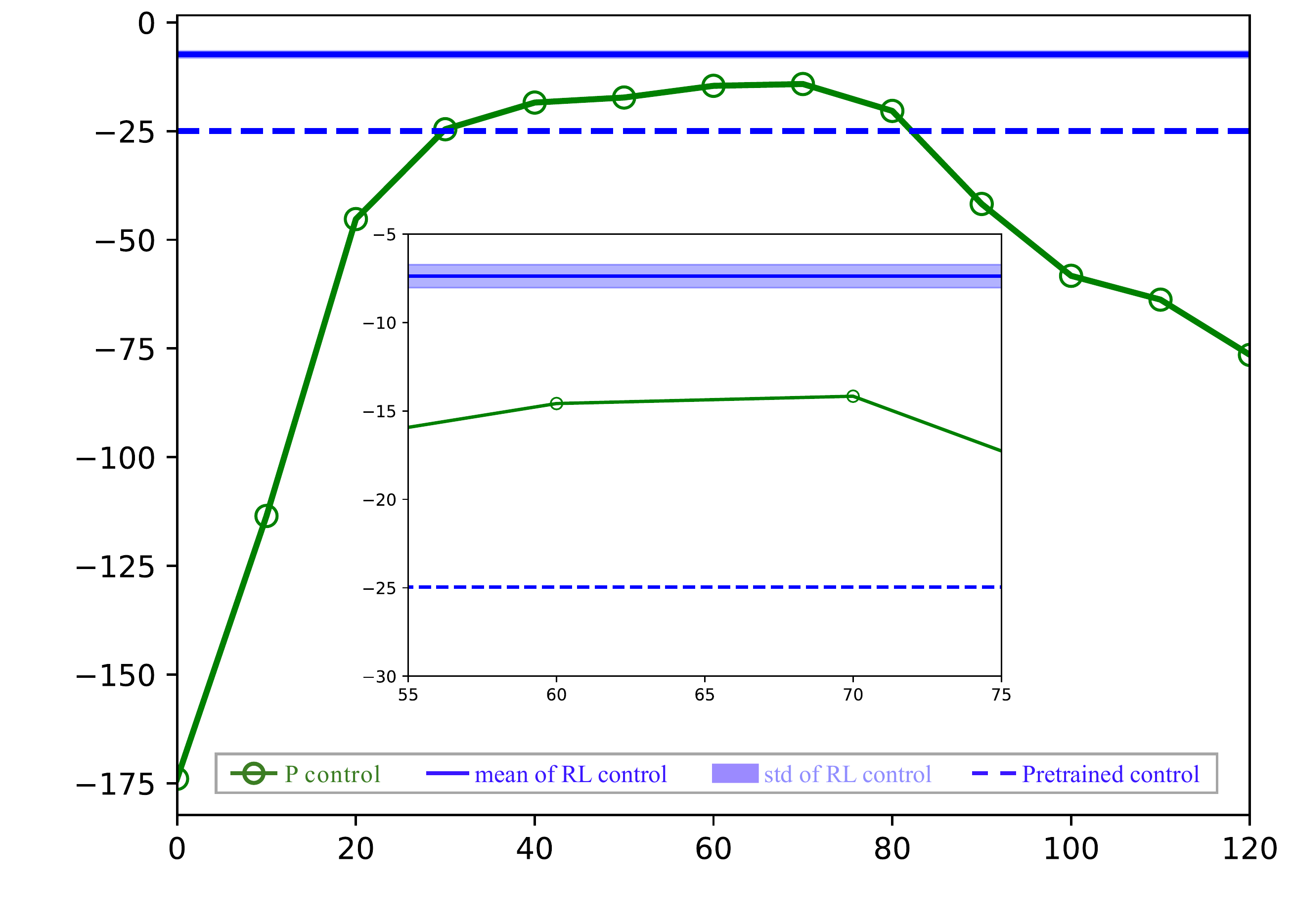}
    \put(2,37){\rotatebox{90}{$\sum r_t$}}
    \put(54,1){{$K_p$}}
    \put(25,34){\rotatebox{90}{$\sum r_t$}}
    \put(53,14.5){{$K_p$}}
    \end{overpic}
    \caption{Evaluation of proportional control and RL control in three-gust environment. Each P control data point is averaged over $45$ random cases. The mean and standard deviation value for the RL agents are calculated from the last $5$ episodes.}
    \label{fig:eval_three_mid}
\end{figure}

In Fig. \ref{fig:eval_three_mid} we evaluate the trained RL control strategies against the P control strategies over $45$ random cases with three gusts. First, we observe that the case with no control yields a total reward of around $-175$, which is significantly lower than the corresponding value of approximately $15$ in a single-gust environment and reflects the stronger disturbance introduced by three successive gust. Furthermore, the total reward increases with the $K_p$ value until it reaches the peak total reward of around $-14$ at $K_p=70$. Beyond this point, the performance starts to deteriorate, likely due to the system becoming overly reactive to the gust. Meanwhile, the RL agent shows a clear improvement in performance from the initial pretrained policy (blue dashed line) to the final pretrained policy (blue envelope). The final RL control surpasses the best P control by a reward margin of approximately $7$. This performance gap is notably larger than the marginal $0.4$ improvement observed in the single-gust environment. \revision{This improvement shows the superiority of the RL controller, compared to the P controller, in a more challenging multi-gust environment. There are at least two plausible reasons for this improvement: first, the ability of the transformer-based RL controller to exploit the temporal information (i.e. the observation history) becomes even more beneficial in such environments; second, the neural networks may provide a richer nonlinear mapping from observations to actions than a proportional mapping, which may also contribute to this observed gain. A more definitive attribution of this improvement would require a dedicated comparison against linear dynamic controllers with comparable temporal dependencies, which we regard as an interesting direction for future work.}

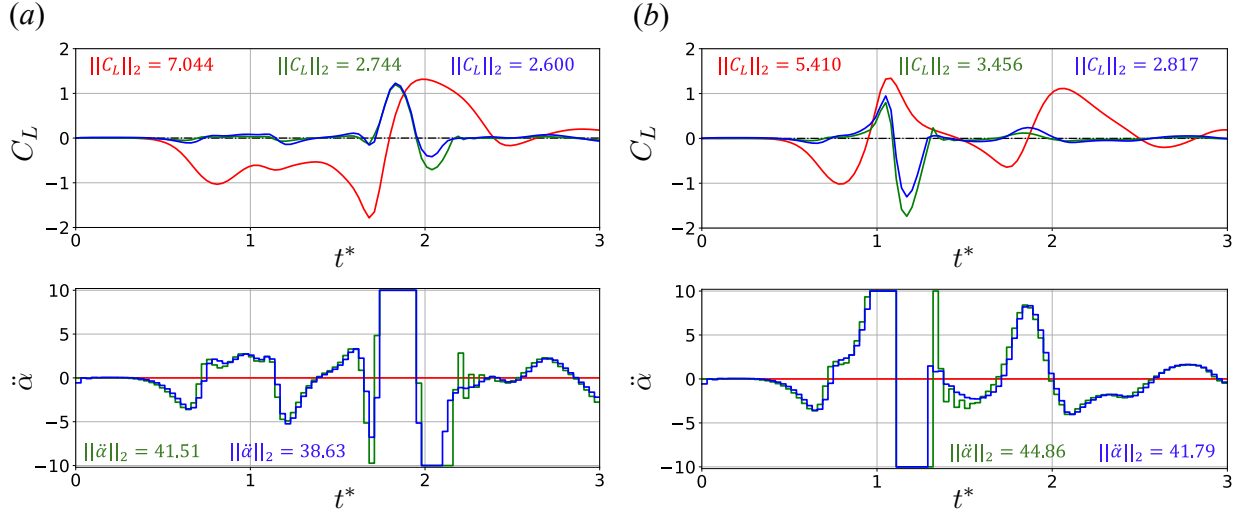
\begin{figure}[htbp]
    \centering
    \begin{overpic}[width=1.0\linewidth]{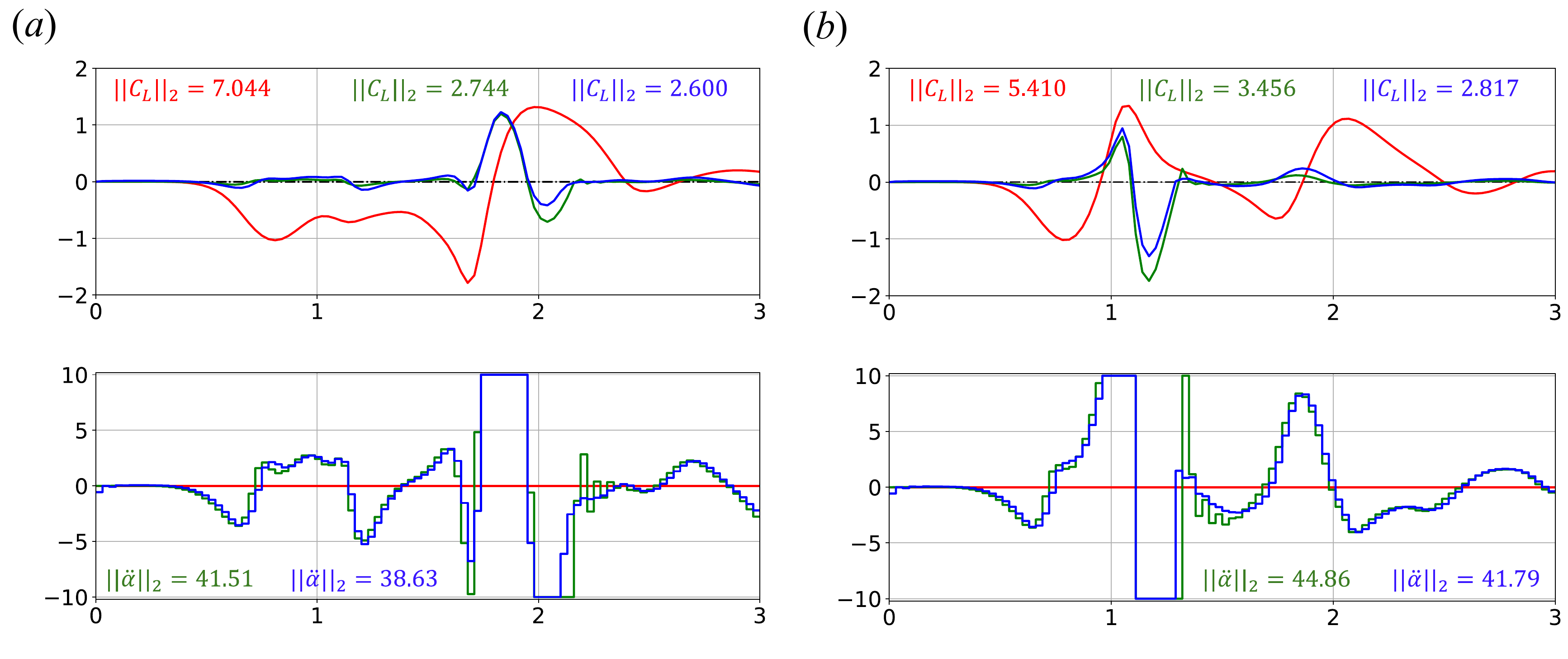}
    \put(1,28.5){\rotatebox{90}{$C_L$}}
    \put(27,20){{$t^*$}}
    \put(51,28.5){\rotatebox{90}{$C_L$}}
    \put(77,20){{$t^*$}}
    \put(1,10){\rotatebox{90}{$\ddot{\alpha}$}}
    \put(27,0.5){{$t^*$}}
    \put(51,10){\rotatebox{90}{$\ddot{\alpha}$}}
    \put(77,0.5){{$t^*$}}
    \end{overpic}
    \caption{The lift histories and angular acceleration histories under no control, P control, and RL control for representative three-gust case 1 in (\textit{a}), and representative single-gust case 2 in (\textit{b}).  The color convention is \textcolor{red}{\rule[0.5ex]{1.5em}{1pt}: no control}, \textcolor{green!60!black}{\rule[0.5ex]{1.5em}{1pt}: P control},
\textcolor{blue}{\rule[0.5ex]{1.5em}{1pt}: RL control}. The corresponding total rewards are: (\textit{a}) \textcolor{red}{$\sum r_t=-451.24$}, \textcolor{green!60!black}{$\sum r_t=4.88$}, \textcolor{blue}{$\sum r_t=15.46$}; (\textit{b}) \textcolor{red}{$\sum r_t=-225.16$}, \textcolor{green!60!black}{$\sum r_t=-43.53$}, \textcolor{blue}{$\sum r_t=3.29$}.}
    \label{fig:lift_compare_three_mid}
\end{figure}

Similar to Section \ref{subsec:benchmarking}, in Fig. \ref{fig:lift_compare_three_mid} we select two representative cases to show the corresponding lift and action histories and report the corresponding norms $||C_L||_2$ and $||\ddot{\alpha}||_2$. The P control corresponds to $K_p=70$ due to its optimal performance in Fig. \ref{fig:eval_three_mid}. The RL control shows a smaller $||C_L||_2$ and $||\ddot{\alpha}||_2$ compared to P control in both of the two representative cases, which means that the RL control mitigates the lift tracking error energy better with less control effort. Both of these facts contribute to the higher total reward for RL control compared to P control. These improvements are also visually evident in the lift histories. For example, in representative case 1, the lift history under RL control shows a smaller disturbance bump compared to the one under P control at around convective time $2$. Similarly, in representative case 2, the RL control results in a smaller lift deviation against P control at convective time approximately $1.2$. These improvements likely stem from subtle refinements in the airfoil kinematics, driven by small but effective modifications in the $\ddot{\alpha}$ profile learned by the RL policy. Despite these improvements, we can still see some notable lift variations under the RL control. These remaining deviations are primarily due to the magnitude limit we impose on the $\ddot{\alpha}$, which constrains the ability of the controller to counteract a strong gust. This can be confirmed by the correspondence between the intervals of the notable lift variation and the saturation of $\ddot{\alpha}$ at the upper or lower bounds. Overall, these findings illustrate that, even when the control authority is limited, the RL control can learn to modulate the actuation to extract better performance.

\begin{figure}[htbp]
    \centering
    \begin{overpic}[width=1.0\linewidth]{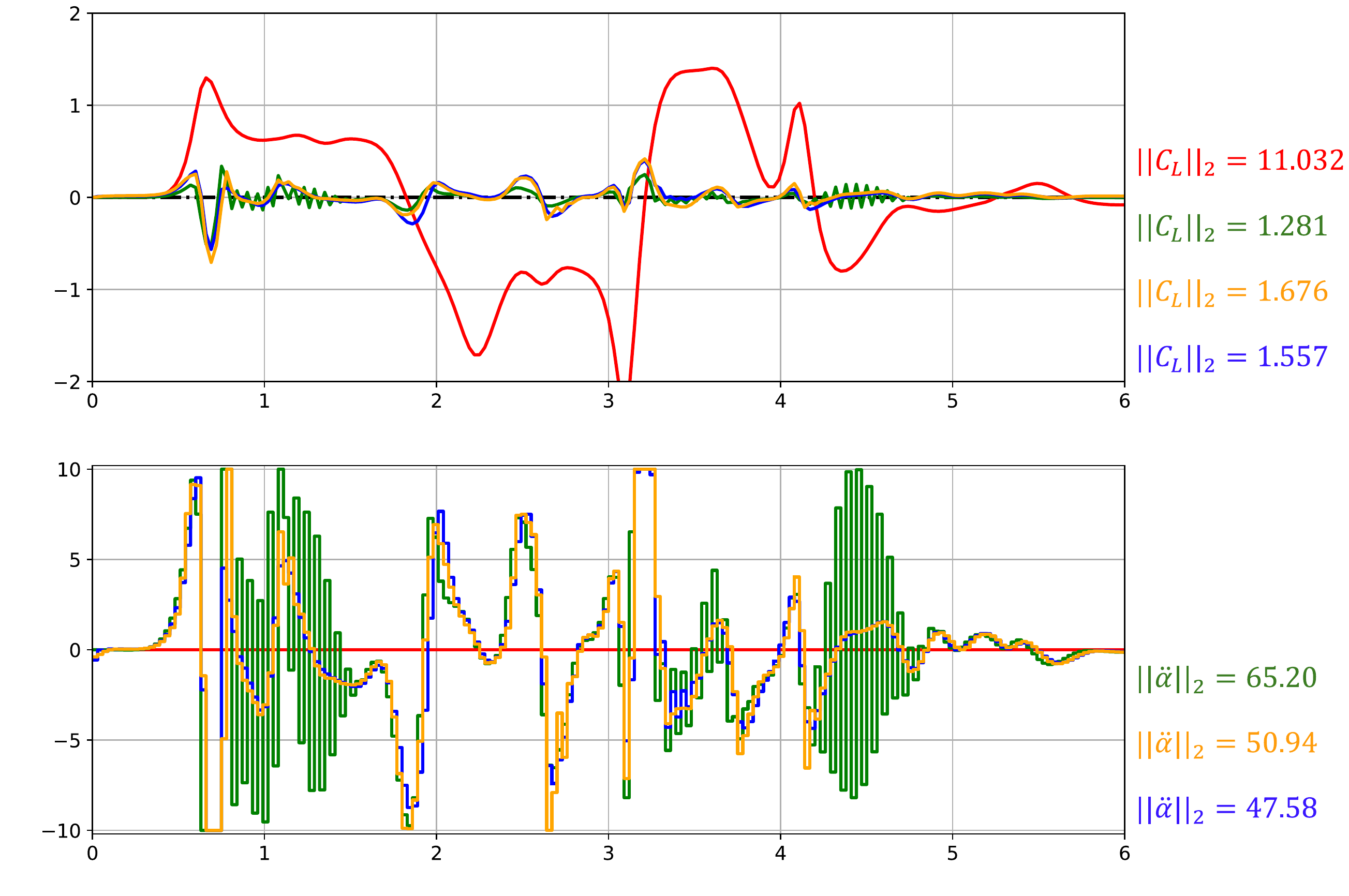}
    \put(1,50){\rotatebox{90}{$C_L$}}
    \put(44,33.5){{$t^*$}}
    \put(1,17){\rotatebox{90}{$\ddot{\alpha}$}}
    \put(44,0){{$t^*$}}
    \end{overpic}
    \caption{The lift histories and angular acceleration histories under no control, P control, RL control trained from single-gust environment, and RL control trained from three-gust environment for one representative case in eight gust environment.  The color convention is \textcolor{red}{\rule[0.5ex]{1.5em}{1pt}: no control}, \textcolor{green!60!black}{\rule[0.5ex]{1.5em}{1pt}: P control}, \textcolor{orange}{\rule[0.5ex]{1.5em}{1pt}: RL control trained in single-gust environment}, \textcolor{blue}{\rule[0.5ex]{1.5em}{1pt}: RL control trained in three-gust environment}. The corresponding total rewards are: \textcolor{red}{$\sum r_t=-1152.17$}, \textcolor{green!60!black}{$\sum r_t=109.17$}, \textcolor{orange}{$\sum r_t=142.84$}, \textcolor{blue}{$\sum r_t=150.77$}.}
    \label{fig:lift_compare_eight_mid}
\end{figure}

We now turn our attention to the eight-gust environment, which approximates the \textit{n}-successive gust environment commonly encountered in real-world applications. Fig. \ref{fig:lift_compare_eight_mid} shows the lift and action histories for a representative eight-gust case under four control strategies: no control, P control, RL control trained in a single-gust environment, and RL control trained in a three-gust environment. Due to the extremely large parameter space associated with the eight-gust setup, performing a comprehensive statistical evaluation---similar to those conducted for the single- and three-gust cases---would require hundreds of simulation runs per control strategy, so it is computationally prohibitive. Therefore, instead of identifying an optimal $K_p$ through exhaustive evaluation, we adopt a more feasible approach. Considering that the optimal gains in single-gust and three-gust environments are both approximately $K_p=70$, we simply choose $K_p=70$ as a reasonable representative value.

Note that the simulation here lasts for $6$ convective time units, comprising 200 control steps, so the ideal total reward is $200$ if $C_L$ remains zero with no control effort throughout the episode. We can first observe from $||C_L||_2$ and $||\ddot{\alpha}||_2$ in Fig. \ref{fig:lift_compare_eight_mid} that both RL agents achieve a comparable lift regulation, slightly larger than the P controller but requiring significantly less control effort. Therefore, the total rewards of the two RL agents are higher than the P controller. Notably, we see that the P control exhibits a strong temporal fluctuation in the $\ddot{\alpha}$ history, and this is also reflected somewhat in the corresponding $C_L$ history. This behavior indicates that the P controller tends to overreact to the successive gust disturbances. In addition, by comparing the two RL agents, we find that the RL agent trained in the three-gust environment shows a slightly better lift regulation with less control effort, but the performance gap between them remains modest. This result is somewhat surprising, as we initially expected that the richer dynamics in the three-gust environment---including nonlinear interactions between successive gusts---would provide more informative training and lead to greater generalization capability. This lack of significant difference may be attributed to the magnitude limit we impose on the $\ddot{\alpha}$. In particular, we observe that the RL agent trained in a single-gust environment performs very well during most of the episode except for the time intervals around $0.7$ and $3.2$, when the lift disturbance remains insufficiently mitigated, and these two time instances both coincide to points when the $\ddot{\alpha}$ is saturated in both controllers. Even with additional training experience, the authority of the mid-chord pitch actuation is limited, and the three-gust RL agent cannot simply apply a stronger control input to suppress the disturbance. Instead, it must improve performance by subtly adjusting preceding control actions to influence the future airfoil kinematics and lift response---a capability that may only provide incremental improvement. Further evidence for the hypothesis will be shown and discussed in Section\ref{subsec:transferring}, in which the $\ddot{\alpha}$ for quarter-chord pitching remains unsaturated throughout the episode. Nonetheless, it is noteworthy that the RL agent trained in a three-gust environment achieves a total reward $\sum r_t=150.77$, so it is reasonable to conclude that the RL agent trained in an environment with a finite number of gust can still be an effective control strategy in an environment with an infinite number of gusts.

In summary, our results confirm that the performance gap between RL and P control widens in the multi-gust environment. Moreover, we observe that both RL agents trained in single- and three-gust environments can be applied to \textit{n}-successive gusts effectively, and that the RL agent trained in three gusts generalizes slightly better than the one trained on a single gust, but the margin of improvement is moderate due to the actuation magnitude limit. These insights suggest that further performance gains may be achievable with increased control authority, and thus motivate us to examine the effect of pivot point location on control effectiveness in the next sections.

\subsection{Revisiting the pivot location: quarter-chord pitching in single-gust environment}
\label{subsec:revisiting}
The previous results highlighted that, while RL control in a mid-chord pitching configuration can effectively mitigate lift variation, its performance may be limited by the imposed $\ddot{\alpha}$ constraint. To explore whether an alternative pivot point location can improve the control performance, we now revisit the pitching configuration and shift the pivot point to the quarter-chord location. This is a classical choice in unsteady aerodynamic flow, so we expect it to provide an effective control strategy. Before we explore the \textit{n}-successive gust scenario, it is prudent to first investigate the single-gust environment.

The lift on an airfoil encountering a gust and pitching about zero angle of incidence can be decomposed into three terms: 
\begin{equation}\label{eq:lift_decomp}
    C_L = C_{L_\amass}+C_{L_\gust}+C_{L_\resp},
\end{equation}
where $C_{L_\amass}$ represents the inertial reaction (i.e., added mass) of the fluid due to airfoil acceleration, $C_{L_\gust}$ represents the contribution from the gust with no control (i.e., on the airfoil at zero angle of attack), and the remaining lift $C_{L_\resp}$ approximately represents the aerodynamic response to the gust due to pitching kinematics (through the kinematics' modification of fluid vorticity).

We define the first of these, the added mass contribution, using potential flow theory \cite{inviscidbook}:
\begin{equation}\label{eq:added_mass}
    C_{L_\amass}=\underbrace{\frac{\pi c d}{2U_{\infty}^2}\ddot{\alpha}\cos\alpha}_{C_{L_{\ddot{\alpha}}}} + \underbrace{\frac{-\pi c}{2U_{\infty}}\dot{\alpha}\cos2\alpha}_{C_{L_{\dot{\alpha}}}},
\end{equation}
where $d$ is the distance between the pivot point and mid-chord point, defined as positive if the pivot is downstream of the mid-chord and negative if it is upstream. For the mid-chord pitching pursued in the previous sections, $C_{L_{\ddot{\alpha}}}=0$ due to $d=0$; in contrast, for the quarter-chord pitching studied in this section, $d=-c/4$. We note again that the pitching kinematics in this study is defined as positive for counter-clockwise rotation. 

Interpreted in this decomposition, our objective for maintaining zero lift is equivalent to controlling the airfoil so that $C_{L_\amass}+C_{L_\resp}$ compensates the effect from the gust, $C_{L_\gust}$. In the previous mid-chord pitching configuration, the only available terms for control are $C_{L_{\dot{\alpha}}}$ and $C_{L_\resp}$. It is important to note that neither of the two terms is directly dependent on the control input $\ddot{\alpha}$, but only on its integral $\dot{\alpha}$, so the controller is somewhat limited in its authority. In contrast, the quarter-chord pitching configuration also provides access to $C_{L_{\ddot{\alpha}}}$ for control, empowering the controller with more flexibility and higher responsiveness by adjusting this term directly. As we will show in subsequent results, $\dot{\alpha}$ is often smaller by an order of magnitude than $\ddot{\alpha}$ under various control strategies, which suggests that the actuation may not be as likely to saturate at the magnitude constraint in the quarter-chord pitching configuration.

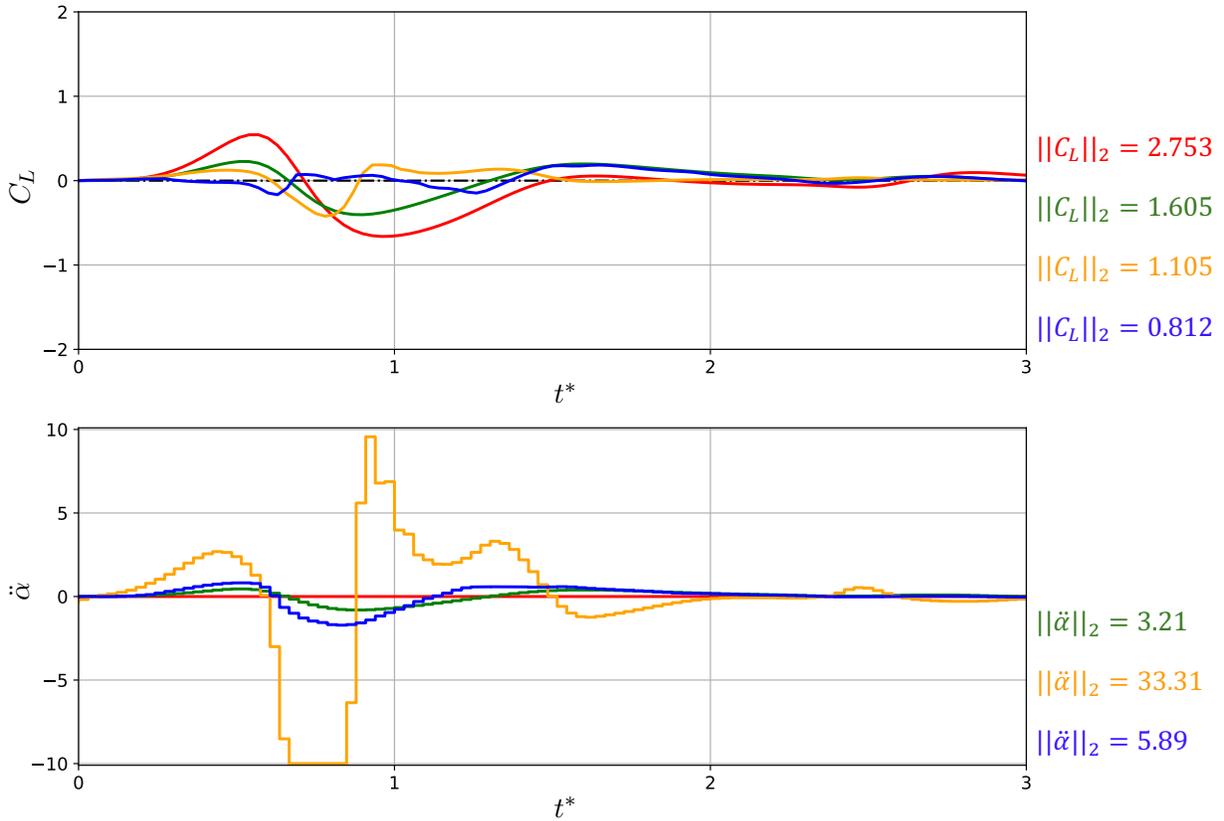
\begin{figure}[htbp]
    \centering
    \begin{overpic}[width=1.0\linewidth]{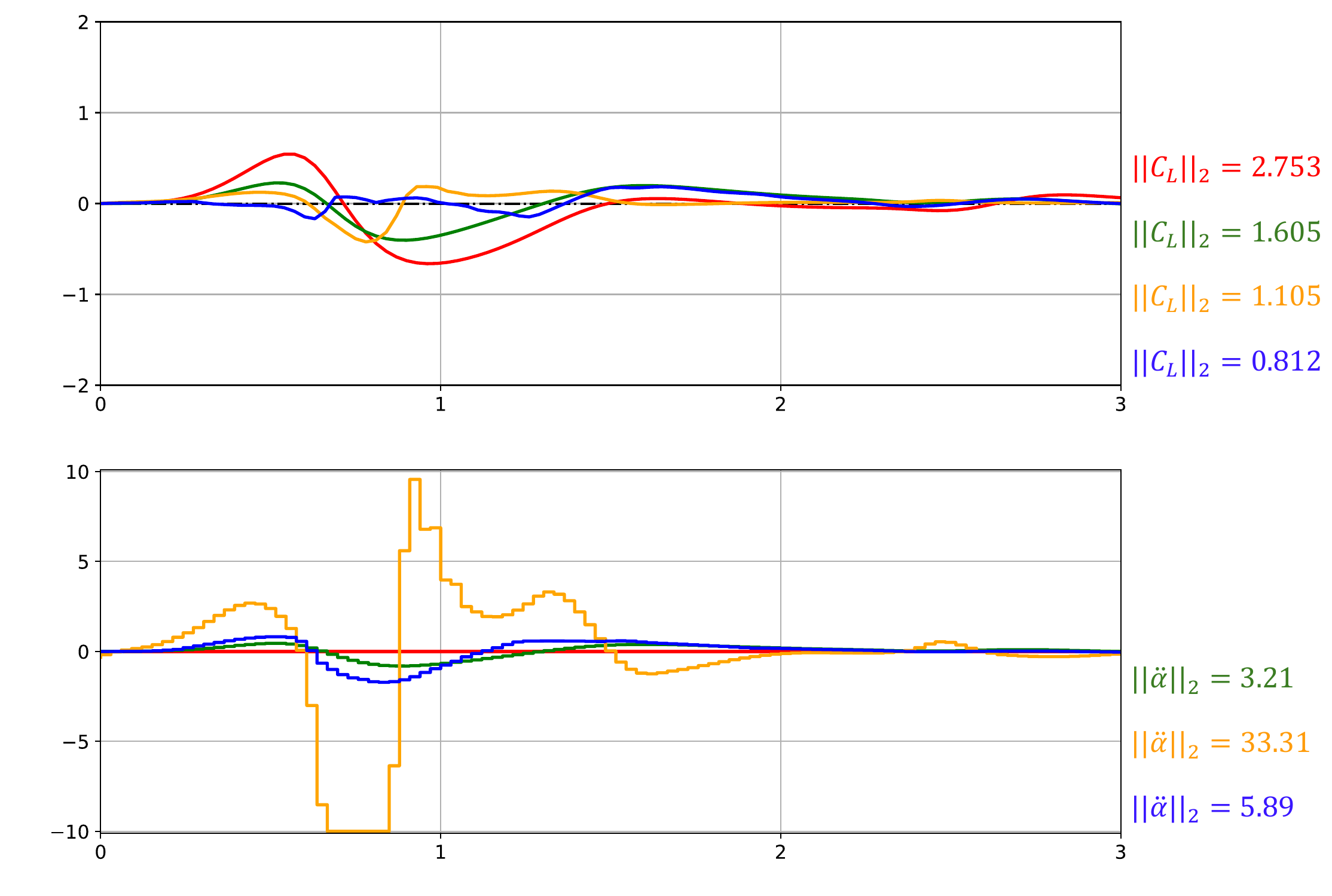}
    \put(2,50){\rotatebox{90}{$C_L$}}
    \put(46,34){{$t^*$}}
    \put(2,18){\rotatebox{90}{$\ddot{\alpha}$}}
    \put(46,0.5){{$t^*$}}
    \end{overpic}
    \caption{The lift histories and angular acceleration histories under no control, quarter-chord P control, mid-chord RL control, and quarter-chord RL control for one representative case in single-gust environment. The color convention is \textcolor{red}{\rule[0.5ex]{1.5em}{1pt}: no control}, \textcolor{green!60!black}{\rule[0.5ex]{1.5em}{1pt}: quarter-chord P control}, \textcolor{orange}{\rule[0.5ex]{1.5em}{1pt}: mid-chord RL control}, \textcolor{blue}{\rule[0.5ex]{1.5em}{1pt}: quarter-chord RL control}. The corresponding total rewards are: \textcolor{red}{$\sum r_t=15.78$}, \textcolor{green!60!black}{$\sum r_t=71.02$}, \textcolor{orange}{$\sum r_t=81.22$}, \textcolor{blue}{$\sum r_t=92.02$}.}
    \label{fig:lift_mid_quarter}
\end{figure}

However, the increased responsiveness also raises a challenge: the P control that is suitable in the mid-chord pitching configuration now may overreact to the lift disturbance in this quarter-chord configuration, and therefore cannot be easily leveraged for pretraining the RL controller. This overreaction is due to the fact that the angular acceleration obtained from the control law \eqref{eq:Pcontrol} is one control step later than the measured lift from which it is determined, while the lift $C_{L_{\ddot{\alpha}}}$ is instantaneously dependent on the angular acceleration. An evaluation of the P control with different gains is provided in Table \ref{tab:p_control}. At large $K_p$, the controller incurs a very large negative reward, indicating exceptionally poor performance that is primarily penalized by the quadratic lift term. At small gain $K_p$ it is possible to maintain a stable lift history, and the total reward $\sum r_t = 74.29$ at $K_p=2$ seems to indicate good performance. However, we will show later in Fig. \ref{fig:lift_mid_quarter} that this high reward is attributable more to the small magnitude of the $\ddot{\alpha}$ rather than successful lift regulation. 

\begin{table}
\caption{For quarter-chord pitching, evaluation of the P control with different gains and RL control over $15$ random cases.}
\label{tab:p_control}
\begin{ruledtabular}
    \begin{tabular}{cccccccc}
      & \makecell{P control \\ $K_p=0$} & \makecell{P control \\ $K_p=1$} & \makecell{P control \\ $K_p=2$} & \makecell{P control \\ $K_p=3$} & \makecell{P control \\ $K_p=4$} & \makecell{RL agents \\ mean} & \makecell{RL agents \\ std}\\
      \hline
      $\sum r_t$ & $16.64$ & $53.71$ & $74.29$   & $-14599.92$ & $-18998.54$ & $90.33$ & $0.64$\\
    \end{tabular}
\end{ruledtabular}
\end{table}

The small $K_p$ controller also leads to expert data with minimal action variation, which, when used as pretraining data, offers little advantage over the default (random) initialization of the RL policy network. Specifically, since we use zero-center initialization (a standard PyTorch initialization \cite{paszke2019pytorch}) for the policy networks, which produces a near-zero output already, and we employ a fixed initial policy standard deviation of $1$, which allows for substantial exploration early in training, the influence of the expert data is diminished during initial policy updates. Therefore, we chose not to use pretraining in this case. To clarify, our decision to omit pretraining in this configuration is not a limitation of the framework, but rather a deliberate design choice driven by the observed ineffectiveness of expert data in this particular setup. As we will demonstrate, even without pretraining, the RL agent is capable of learning an effective control strategy in the quarter-chord configuration. For the readers who may be interested in pursuing pretraining under these conditions, one potential solution is to develop a full PID controller, which would likely be more successful than P control in reducing the overreaction. One could also apply a scaling factor to amplify the action magnitudes from the expert data before using them for pretraining, which would ensure that the pretrained policy deviates more meaningfully from the zero-center initialized baseline. Another approach may be to employ a smaller initial policy standard deviation, which would constrain the exploration range and help maintain the influence of the pretrained expert data. However, these approaches require careful tuning, and are outside the scope of the present study.

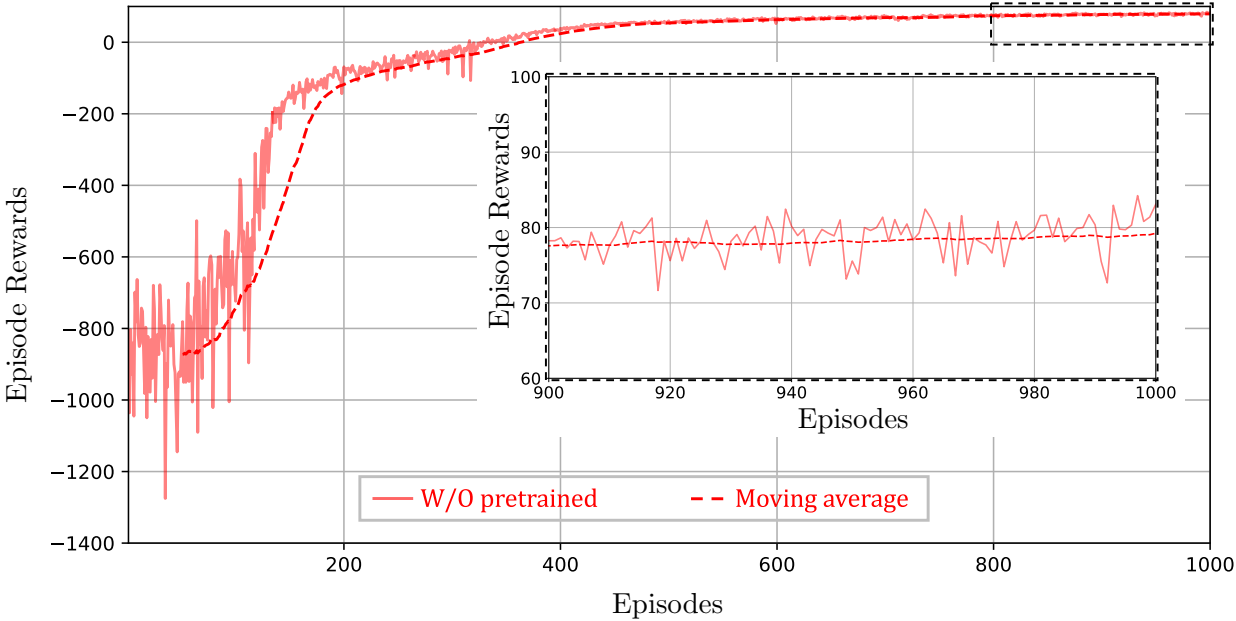
\begin{figure}[htbp]
    \centering
    \begin{overpic}[width=1.0\linewidth]{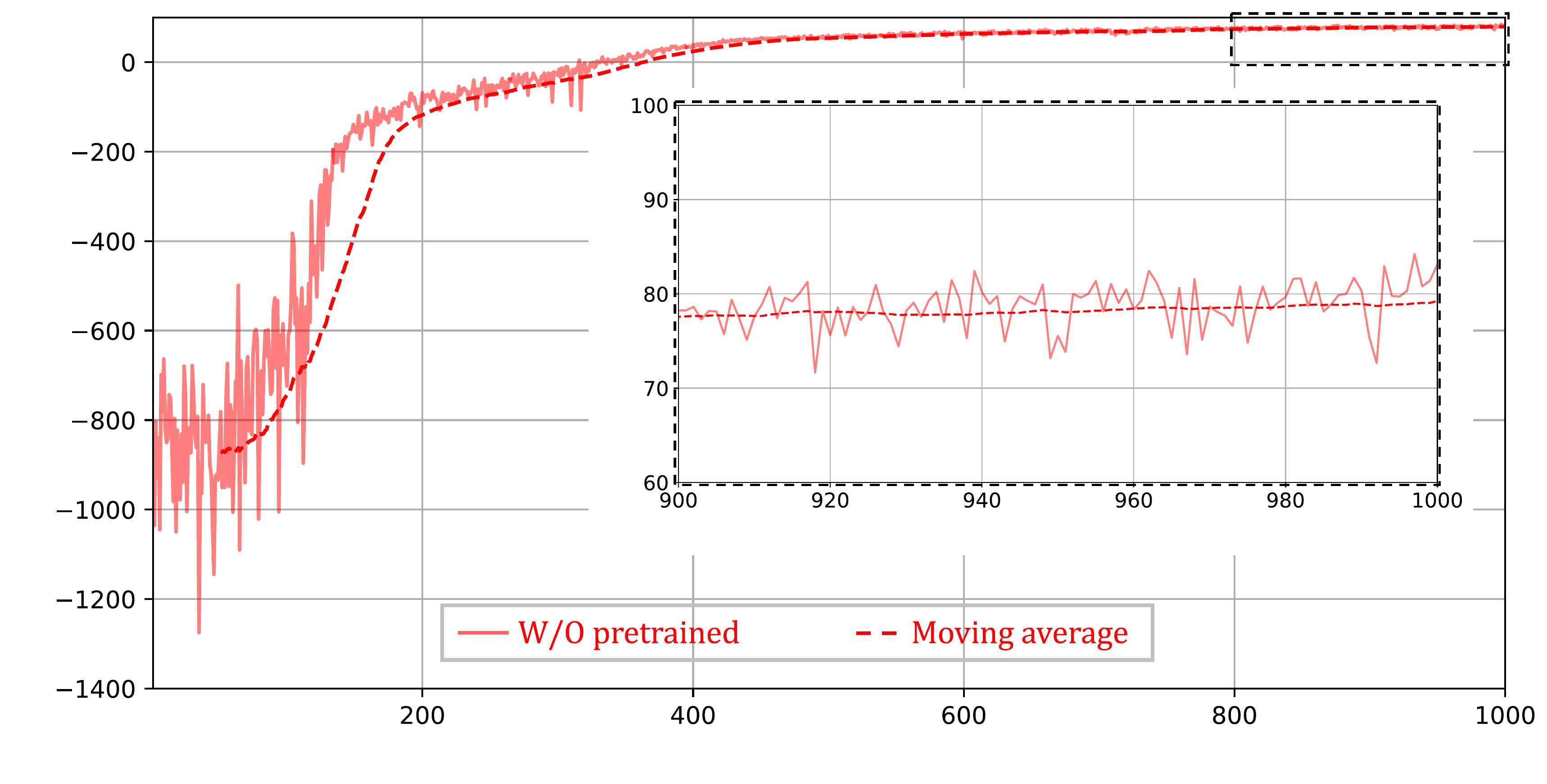}
    \put(0,17.5){\rotatebox{90}{Episode Rewards}}
    \put(49,0.5){{Episodes}}
    \put(39,23.5){\rotatebox{90}{\small Episode Rewards}}
    \put(64,15.5){{\small Episodes}}
    \end{overpic}
    \caption{Learning curve for quarter-chord pitching RL agents in single-gust environment. The solid line represents the history of episode rewards for single run. The dashed line is the moving average with a window size of $50$. At each episode index, the episode reward is averaged across all parallel environments.}
    \label{fig:learning_quarter_single}
\end{figure}

The learning curve for the RL agent in the quarter-chord pitching configuration is depicted in Fig. \ref{fig:learning_quarter_single}. We observe that the learning process starts from an initial reward around $-1000$, which is much smaller than $-430$ in the mid-chord pitching from Fig. \ref{fig:learning_mid_single}. This reflects the unstable lift profile caused by the added-mass term $C_{L_{\ddot{\alpha}}}$ during the early stages of the training. As training proceeds, the reward grows steadily with a decreasing variance, and reaches $0$ at approximately the $300\text{th}$ episode. Beyond this point, the learning proceeds relatively more slowly, and finally reaches around $80$. While the learning curve suggests that the policy may continue to improve marginally, we opt to stop training at this stage to avoid unnecessary computational cost, as the remaining improvements are likely negligible. Based on the final performance of around $80$, which is much larger than $60$ for the performance of mid-chord RL control shown in Fig. \ref{fig:learning_mid_single}, we can already conclude that the quarter-chord RL control is superior.

The evaluation of the performance of the RL agents from the last $5$ episodes is illustrated in Table \ref{tab:p_control}. The RL agents outperform the best P control with a performance gap of around $16$, which is much larger than the gap observed in the mid-chord pitching configuration. To provide further insight, Fig. \ref{fig:lift_mid_quarter} presents the lift and action histories for one representative case. For clarity, the P control shown corresponds to $K_p=2$, which yields the highest performance among the tested gains. From the report of $||C_L||_2$ and $||\ddot{\alpha}||_2$, we can first see that quarter-chord pitching RL control regulates lift much better than quarter-chord P control, and with only slightly more control effort. Consequently, the quarter-chord RL control has a much higher total reward. Furthermore, a comparison between the mid-chord and quarter-chord RL controllers highlights the benefits of the pivot point shift. We can see that the quarter-chord pitching results in a better lift tracking with significantly less control effort. The lift bump under mid-chord RL control at convective time around $0.75$ is absent in the profile under quarter-chord RL control. This aligns with our earlier discussion that the presence of the $C_{L_{\ddot{\alpha}}}$ term in the quarter-chord setup grants the controller direct authority over lift regulation via angular acceleration. As a result, the quarter-chord RL controller does not encounter saturation of $\ddot{\alpha}$, avoiding the performance limitations seen in the mid-chord case.

\begin{figure}[htbp]
    \centering
    \begin{overpic}[width=1.0\linewidth]{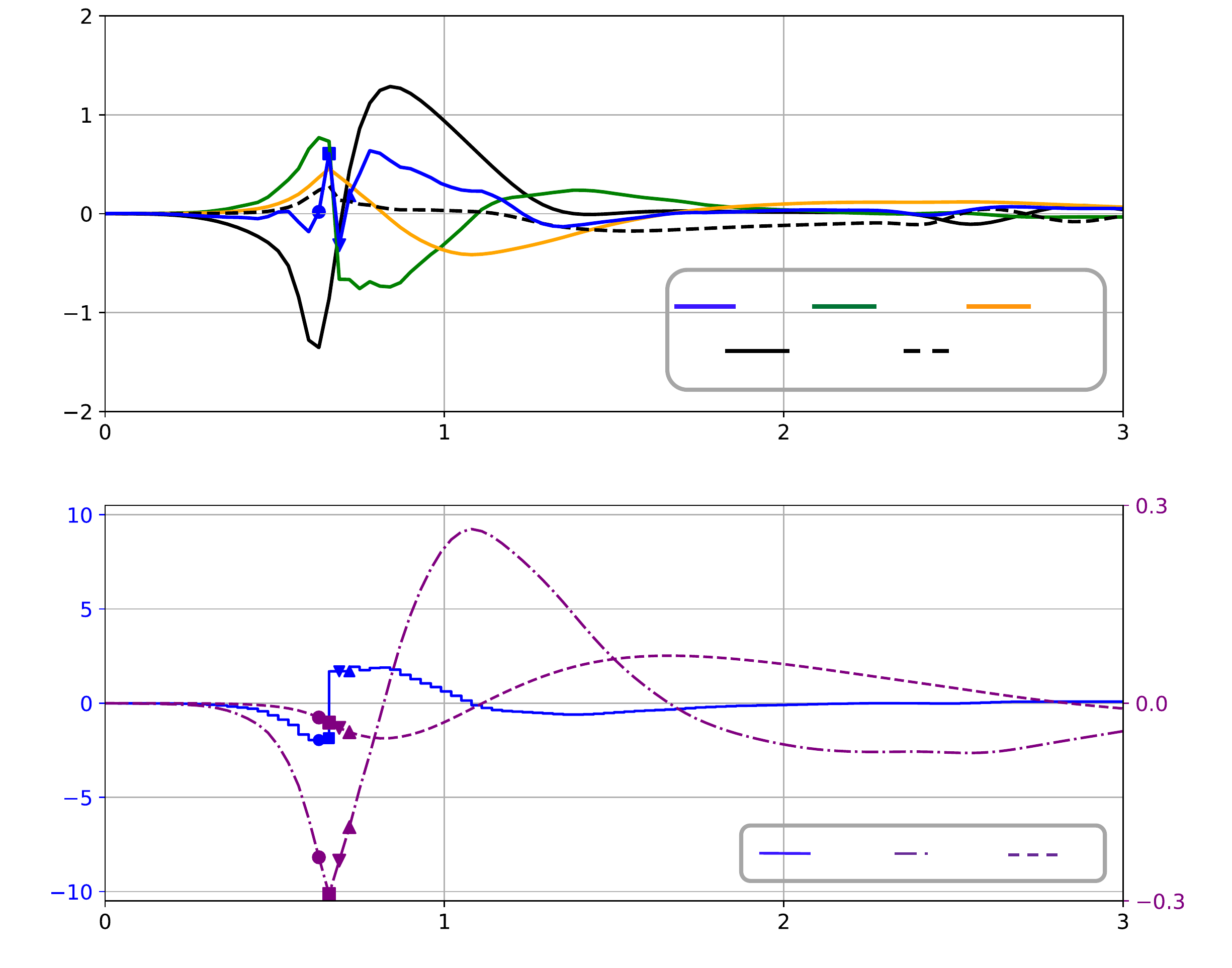}
    \put(2,59.5){\rotatebox{90}{$C_L$}}
    \put(49.5,40){{$t^*$}}
    \put(2,20.5){\rotatebox{90}{$\ddot{\alpha}$}}
    \put(96.5,19.5){\rotatebox{90}{$\alpha,\dot{\alpha}$}}
    \put(49.5,0.5){{$t^*$}}
    \put(60.2,53){{$C_L$}}
    \put(71.5,53){{$C_{L_{\ddot{\alpha}}}$}}
    \put(84.1,53){{$C_{L_{\dot{\alpha}}}$}}
    \put(64.5,49.5){{$C_{L_\gust}$}}
    \put(77.5,49.5){{$C_{L_\resp}$}}
    \put(66,8.5){{$\ddot{\alpha}$}}
    \put(76,8,5){{$\dot{\alpha}$}}
    \put(86.5,8.5){{$\alpha$}}
    \end{overpic}
    \caption{Lift decomposition for one representative case in quarter-chord pitching configuration.}
    \label{fig:lift_decomp_quarter}
\end{figure}

\begin{figure}[htbp]
    \centering
    \includegraphics[width=1.0\linewidth]{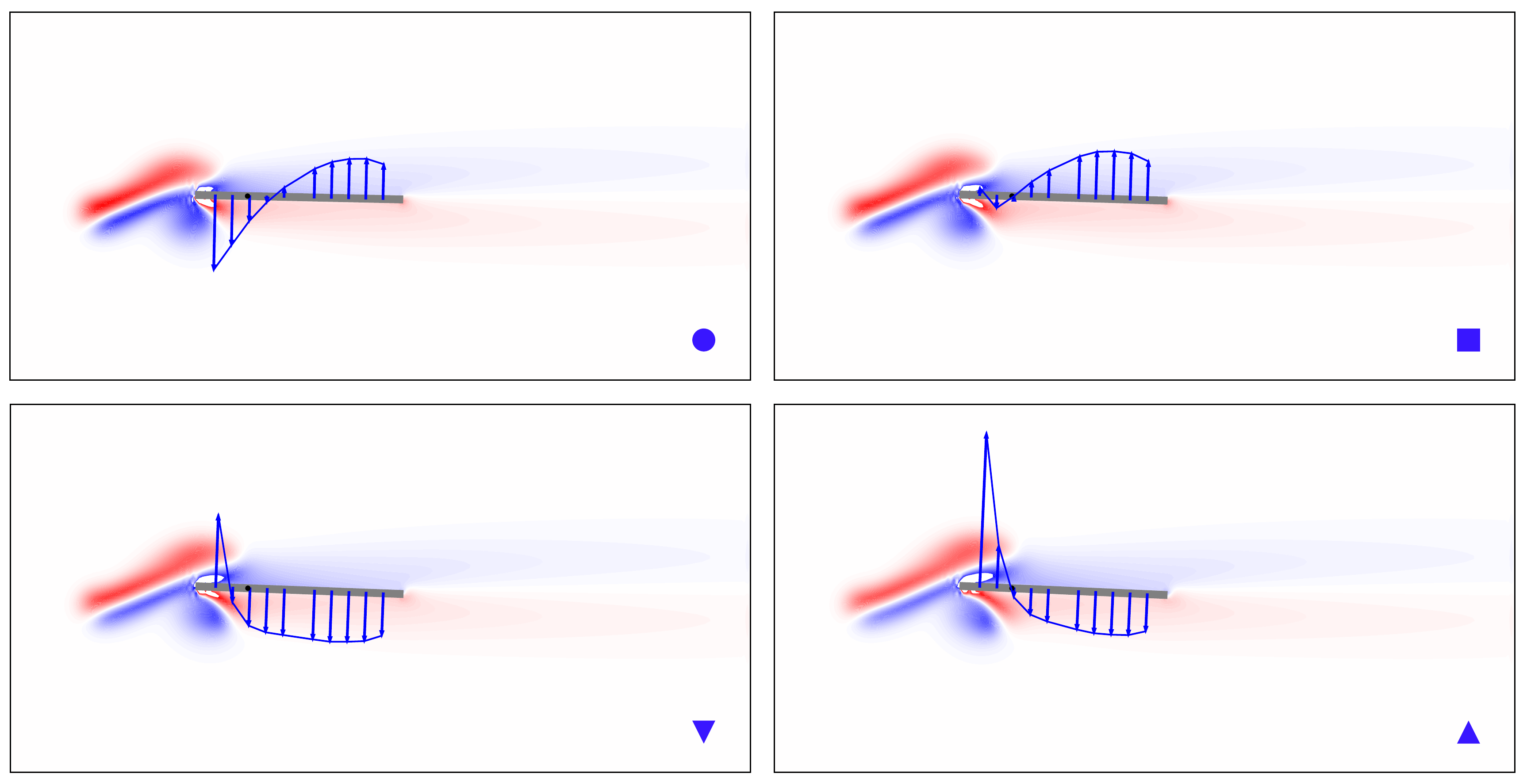}
    \caption{The corresponding vorticity contour and surface pressure distribution at the four steps indicated in Fig. \ref{fig:lift_decomp_quarter}. \revision{Pressure arrows pointing up represent positive $C_{\Delta p}$, and pointing down represent negative $C_{\Delta p}$.}}
    \label{fig:vorticity_quarter_jump}
\end{figure}

\begin{figure}[htbp]
    \centering
    \begin{overpic}[width=1.0\linewidth]{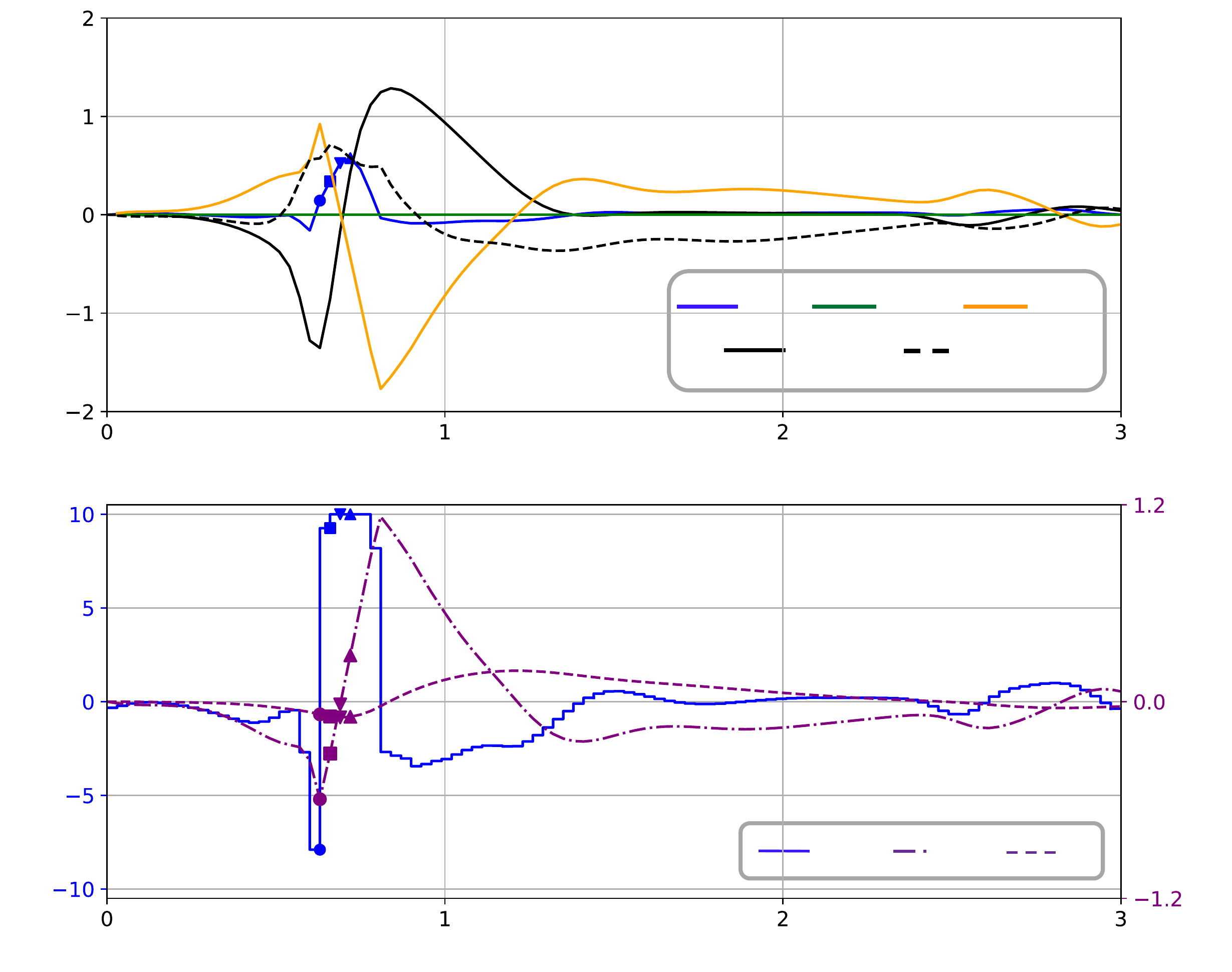}
    \put(2,59.5){\rotatebox{90}{$C_L$}}
    \put(49.5,40){{$t^*$}}
    \put(2,20.5){\rotatebox{90}{$\ddot{\alpha}$}}
    \put(96.5,19.5){\rotatebox{90}{$\alpha,\dot{\alpha}$}}
    \put(49.5,0.5){{$t^*$}}
    \put(60.5,53.2){{$C_L$}}
    \put(71.5,53.2){{$C_{L_{\ddot{\alpha}}}$}}
    \put(83.9,53.2){{$C_{L_{\dot{\alpha}}}$}}
    \put(64.2,49.7){{$C_{L_\gust}$}}
    \put(77.5,49.7){{$C_{L_\resp}$}}
    \put(66,9){{$\ddot{\alpha}$}}
    \put(76,9){{$\dot{\alpha}$}}
    \put(86.3,9){{$\alpha$}}
    \end{overpic}
    \caption{Lift decomposition and the corresponding action history for one representative case in mid-chord pitching configuration.}
    \label{fig:lift_decomp_mid}
\end{figure}

\begin{figure}[htbp]
    \centering
    \includegraphics[width=1.0\linewidth]{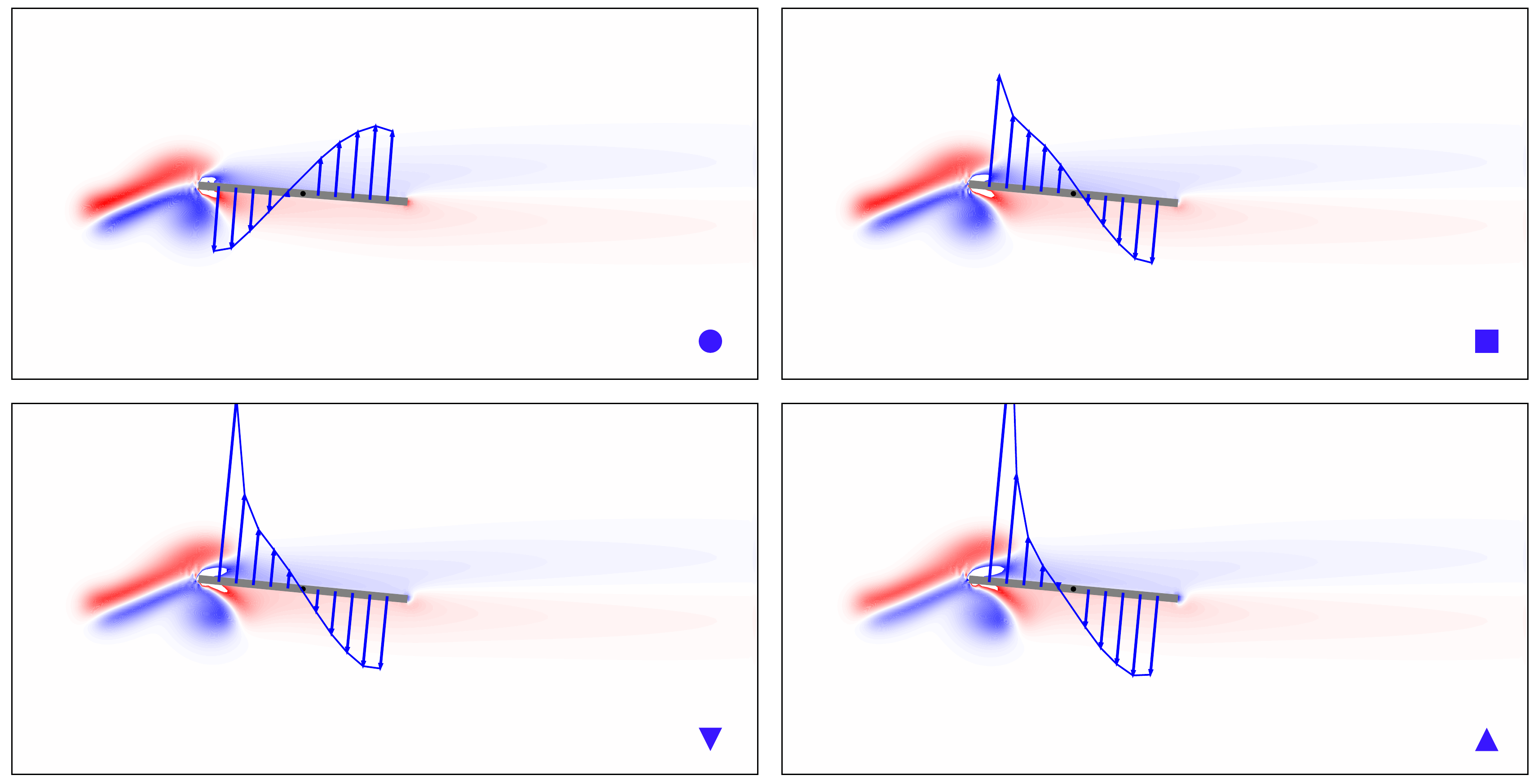}
    \caption{The corresponding vorticity contour and surface pressure distribution at the four steps indicated in Fig. \ref{fig:lift_decomp_mid}. \revision{Pressure arrows pointing up represent positive $C_{\Delta p}$, and pointing down represent negative $C_{\Delta p}$.}}
    \label{fig:vorticity_mid_jump}
\end{figure}

For more specific comparison between the mid-chord and quarter-chord RL control on the basis of flow physics, we present the decomposed lift components based on equations \eqref{eq:lift_decomp} and \eqref{eq:added_mass}, along with the corresponding pitching kinematics histories, for an identical representative case. The results for the quarter-chord and mid-chord configurations are shown in Figs. \ref{fig:lift_decomp_quarter} and \ref{fig:lift_decomp_mid}, respectively. The corresponding vorticity contours and surface pressure distribution for the two configurations at specific time instances ($t^* = 0.6$, $0.63$, $0.66$, $0.69$), which correspond to four consecutive control steps, are shown in Figs. \ref{fig:vorticity_quarter_jump} and \ref{fig:vorticity_mid_jump}, respectively.


From the comparison between Figs. \ref{fig:lift_decomp_quarter} and \ref{fig:lift_decomp_mid}, we can first observe that the quarter-chord RL control can leverage $C_{L_{\ddot{\alpha}}}$ to compensate for $C_{L_{\gust}}$, so that $C_{L_{\dot{\alpha}}}$ and $C_{L_{\resp}}$ magnitudes are relatively smaller than they are for mid-chord pitching. This can also be observed from the magnitudes of the $\dot{\alpha}$ and $\alpha$ in the quarter-chord pitching configuration: both quantities' magnitudes are bounded by $0.3$, much smaller than $1.2$ in the mid-chord pitching configuration. The smaller magnitudes of $\dot{\alpha}$ and $\alpha$ directly correspond to smaller $C_{L_{\dot{\alpha}}}$ in quarter-chord pitching. They also reveal the reason why $C_{L_{\resp}}$ remains small, since a smaller change of pitching kinematics leads to a smaller modification to the vorticity distribution compared to the uncontrolled flow. The events in the interval around $t^*=0.6$ provide a specific illustration of these differences. We notice that both RL agents are pitching up (clockwise angular velocity) during this interval, but the mid-chord RL agent opts to decelerate beyond $t^*=0.6$, while the quarter-chord RL agent chooses to decelerate one step later, i.e., at $t^*=0.63$. This difference is due to the fact that the mid-chord RL agent needs more time to compensate for the gust, $C_{L_{\gust}}$, by modifying $C_{L_{\dot{\alpha}}}$ and $C_{L_{\resp}}$ via $\dot{\alpha}$ and $\alpha$, which are integrals of the action (the angular acceleration). In contrast, the quarter-chord RL agent is less restricted, since it can react through $C_{L_{\ddot{\alpha}}}$, and can do so with a discontinuous change in angular acceleration. In summary, in the quarter-chord pitching configuration, the $C_{L_{\amass}}$ plays the dominant role for compensating for $C_{L_{\gust}}$, but $C_{L_{\amass}}$ and the $C_{L_{\resp}}$ are equivalently important (and less responsive) in mid-chord pitching configuration.

By comparing the pressure and vorticity for the controllers in Figs. \ref{fig:vorticity_quarter_jump} and \ref{fig:vorticity_mid_jump}, we find that both RL agents effectively learn to counteract the effect of the gust on the pressure near the leading edge by leveraging pitching to modify the overall surface pressure. However, they show important differences in how they achieve this. In the mid-chord configuration, the RL agent reverses the direction of angular acceleration $\ddot{\alpha}$ while the leftmost pressure sensor still reads a negative value, indicating a more anticipatory control behavior. In contrast, the quarter-chord RL agent initiates the change in $\ddot{\alpha}$ only after the leftmost pressure sensor signal turns positive. This behavioral difference may be attributed to the additional control authority provided by the $C_{L_{\ddot{\alpha}}}$ term in the quarter-chord configuration as we discussed above, allowing the agent to respond more reactively without requiring as much lead time. However, since the input to the transformer-based policy network consists of a temporal window of past observations, a more detailed investigation than we present in this paper is needed to fully understand how the pressure history is being mapped to the angular acceleration output, and how this mapping differs across pivot configurations.

\subsection{Transferring control strategies: quarter-chord pitching in multi-gust environment}
\label{subsec:transferring}
Having established an effective control strategy for the quarter-chord pitching configuration in the single-gust environment, we now extend our investigation to a more realistic and challenging scenario involving a \textit{n}-successive gust. As we have discussed in Section \ref{subsec:revisiting}, the P control cannot provide us with suitable expert data for pretraining to accelerate the RL training process in the quarter-chord pitching configuration. However, training from scratch in the single-gust environment required $\sim 1000$ episodes, as shown in Fig. \ref{fig:learning_quarter_single}. Therefore, it is computationally infeasible to train the RL agent from scratch in a more complex environment with multiple gusts, as it would require even more episodes to converge. In order to overcome this difficulty, we leverage transfer learning to make the training feasible. Specifically, we treat the previous single-gust environment as the source task, where the RL agent has acquired useful knowledge about the fluid dynamics and control principles. This knowledge is then transferred to the multi-gust environment---our target task---serving as a warm start that accelerates learning and facilitates adaptation to the more complex setting.

\begin{figure}[htbp]
    \centering
    \begin{overpic}[width=1.0\linewidth]{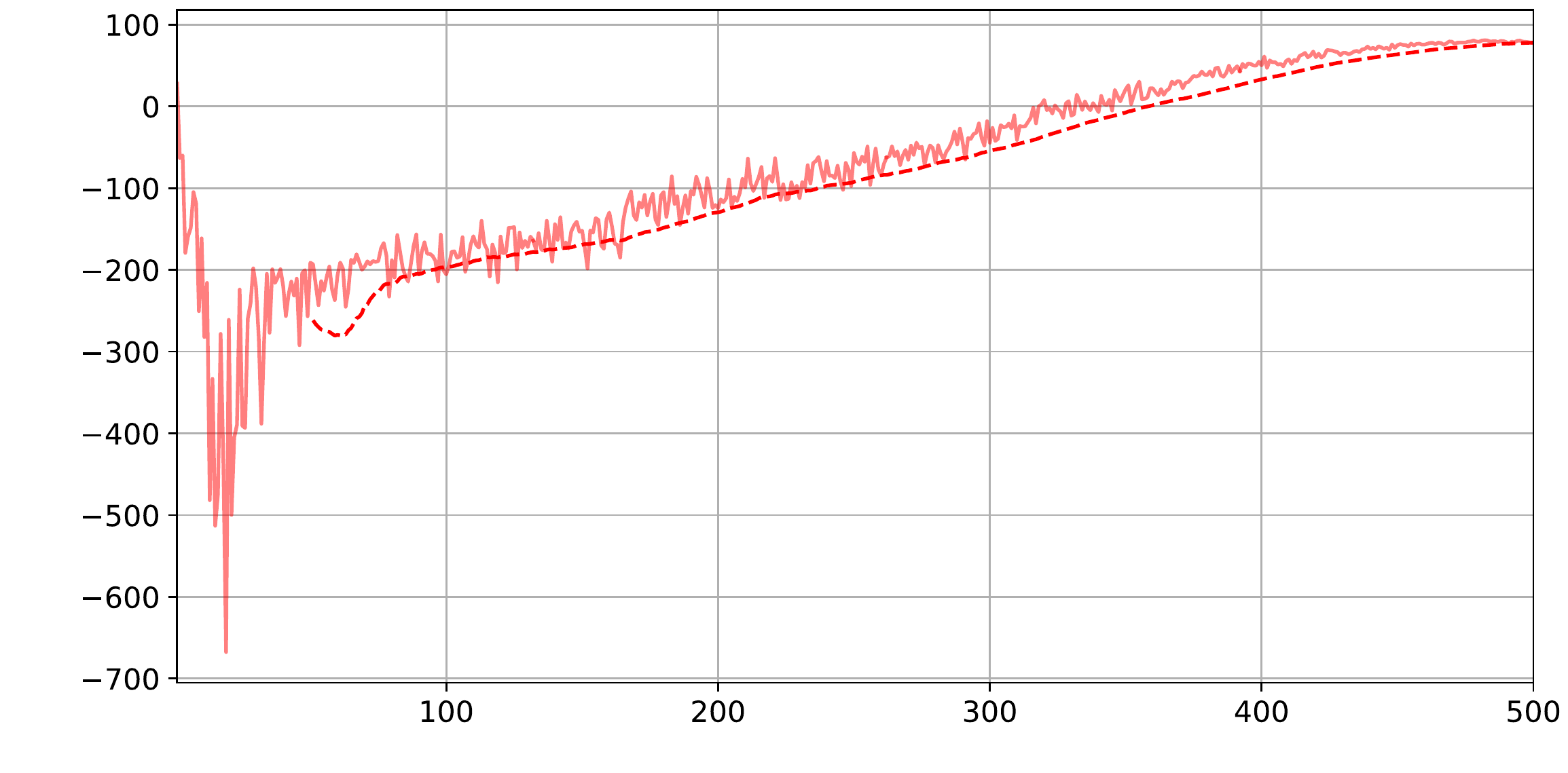}
    \put(1.5,17.5){\rotatebox{90}{Episode Rewards}}
    \put(49.5,0.5){{Episodes}}
    \end{overpic}
    \caption{Learning curve for quarter-chord pitching RL agents in three-gust environment. The solid line represents the history of episode rewards for single run. For the dashed lines, the moving average has a window size of $50$. At each episode index, the episode reward is averaged across all parallel environments.}
    \label{fig:learning_quarter_three}
\end{figure}

The learning curve for the transferred RL agent in a three-gust environment is shown in Fig. \ref{fig:learning_quarter_three}. The initial reward is around $20$, which indicates that the RL agent trained on a single gust generalizes moderately well to the more complex task. A sharp drop occurs in the first $20$ episodes, but then recovers steadily. This sudden drop is characteristic of transfer learning in actor–critic architectures: the policy network may carry over useful features from the source task, but the value network often suffers from inaccurate approximation due to the environment shift between tasks. As a result, early updates primarily focus on correcting the value approximation, rather than improving the control policy, since a large value loss may dominate the total loss function (see equation \eqref{eq:loss}). After this correction phase, the agent resumes improving the policy, surpassing its initial performance at around episode 350 and eventually reaching a final reward close to 80, which is comparable to the final performance in the single-gust environment. This comparable performance is surprising given the increased task complexity. We hypothesize two reasons: First, the richer gust structure in the three-gust environment provides the RL agent with more diverse training signals, potentially enabling it to learn more generalizable or refined control strategies than what the single-gust setup affords. Second, the quarter-chord configuration allows sufficient actuation authority such that the angular acceleration constraint does not limit control performance. Consequently, the RL agent retains the capability to mitigate the stronger disturbance introduced by three gusts without degradation in performance. We note that the learning curve in Fig.\ref{fig:learning_quarter_three} is not yet fully converged, but we choose to stop the training here, anticipating that the subsequent improvement will be small. Compared to the $\sim 1000$ episodes required to train the RL agent from scratch in a single-gust environment, training in this more complex environment with transfer learning only takes $\sim 500$ episodes. Given the increased task complexity, training from scratch in the multi-gust setting would likely require substantially more than $1000$ episodes to reach comparable performance. Therefore, this result demonstrates that transfer learning effectively accelerates the training process and leads to substantial savings in computational resources.

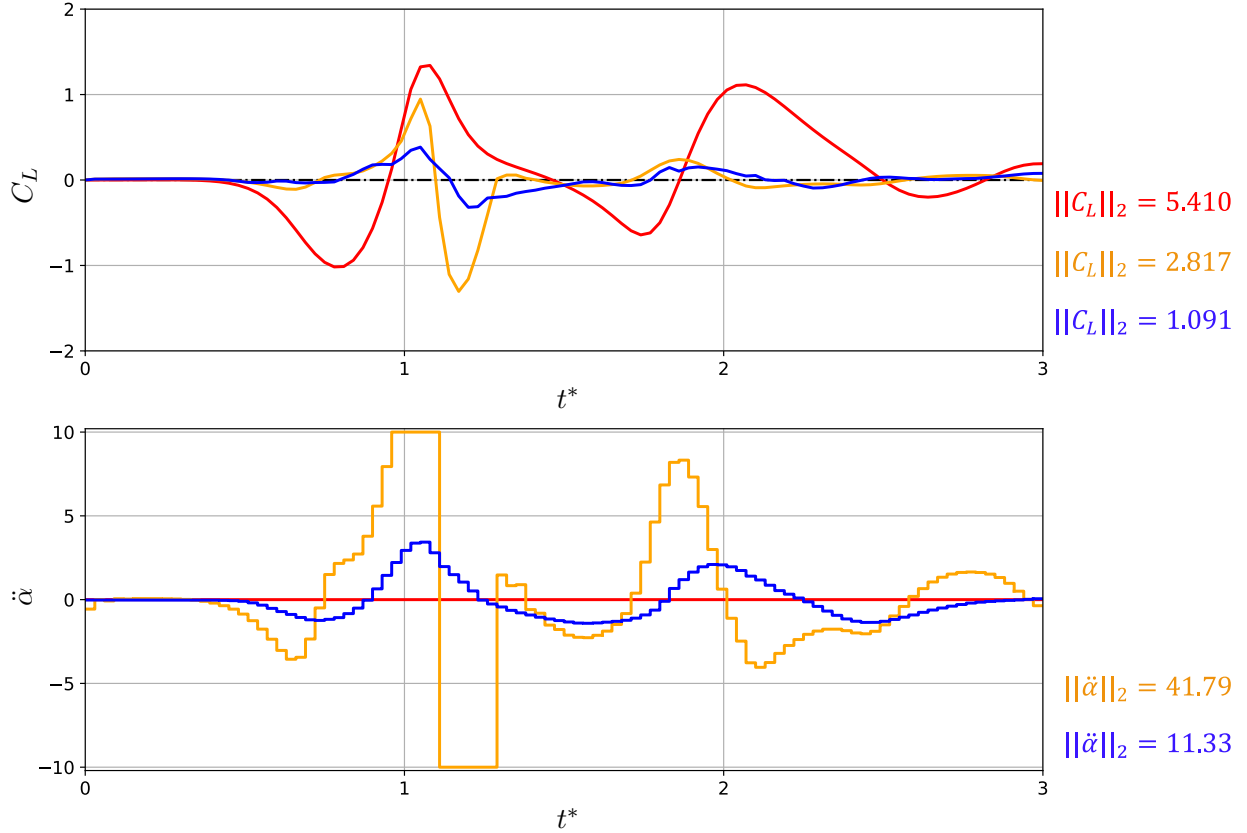
\begin{figure}[htbp]
    \centering
    \begin{overpic}[width=1.0\linewidth]{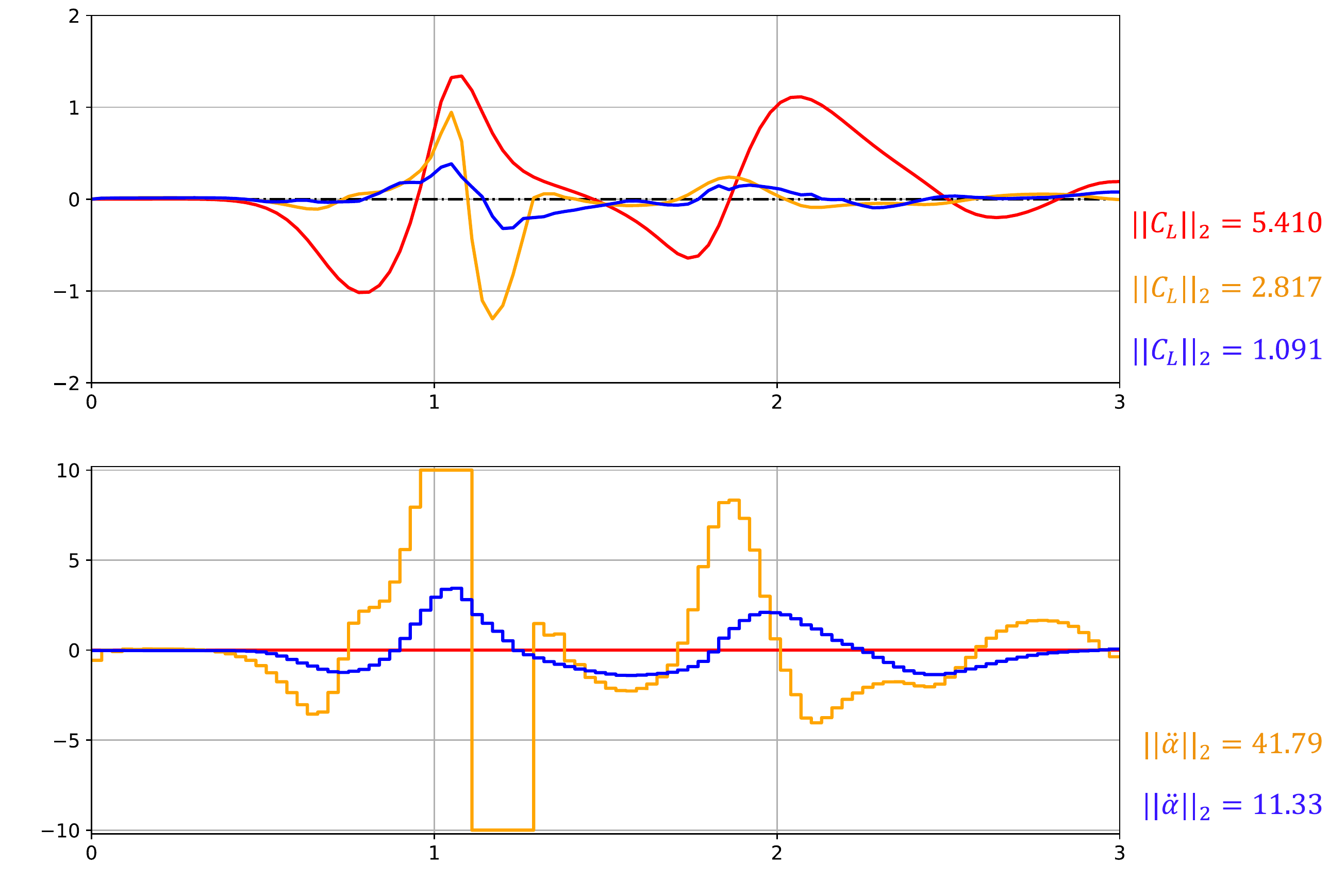}
    \put(1,50.5){\rotatebox{90}{$C_L$}}
    \put(45,34){{$t^*$}}
    \put(1,18){\rotatebox{90}{$\ddot{\alpha}$}}
    \put(45,0){{$t^*$}}
    \end{overpic}
    \caption{The lift histories and angular acceleration histories under no control, mid-chord RL control, and quarter-chord RL control for one representative case in three-gust environment. The color convention is \textcolor{red}{\rule[0.5ex]{1.5em}{1pt}: no control}, \textcolor{orange}{\rule[0.5ex]{1.5em}{1pt}: mid-chord RL control}, \textcolor{blue}{\rule[0.5ex]{1.5em}{1pt}: quarter-chord RL control}. The corresponding total rewards are: \textcolor{red}{$\sum r_t=-225.16$}, \textcolor{orange}{$\sum r_t=3.29$}, \textcolor{blue}{$\sum r_t=84.86$}.}
    \label{fig:lift_mid_quarter_three}
\end{figure}

After the training process, we now evaluate and compare the RL agents in the two different pivot configurations on a representative case in the three-gust environment in Fig. \ref{fig:lift_mid_quarter_three}. From the $||C_L||_2$ and $||\ddot{\alpha}||_2$ comparison, quarter-chord RL control regulates lift better than mid-chord RL control, and with significantly less control effort. Therefore, the total reward for quarter-chord RL control is much higher than the one for mid-chord RL control. Additionally, the lift bumps under mid-chord RL control due to the actuation constraint almost disappear in the lift history under quarter-chord RL control, verifying that the quarter-chord pivot configuration offers the RL agent more control authority even in the more complex environment.

\begin{figure}[htbp]
    \centering
    \begin{overpic}[width=1.0\linewidth]{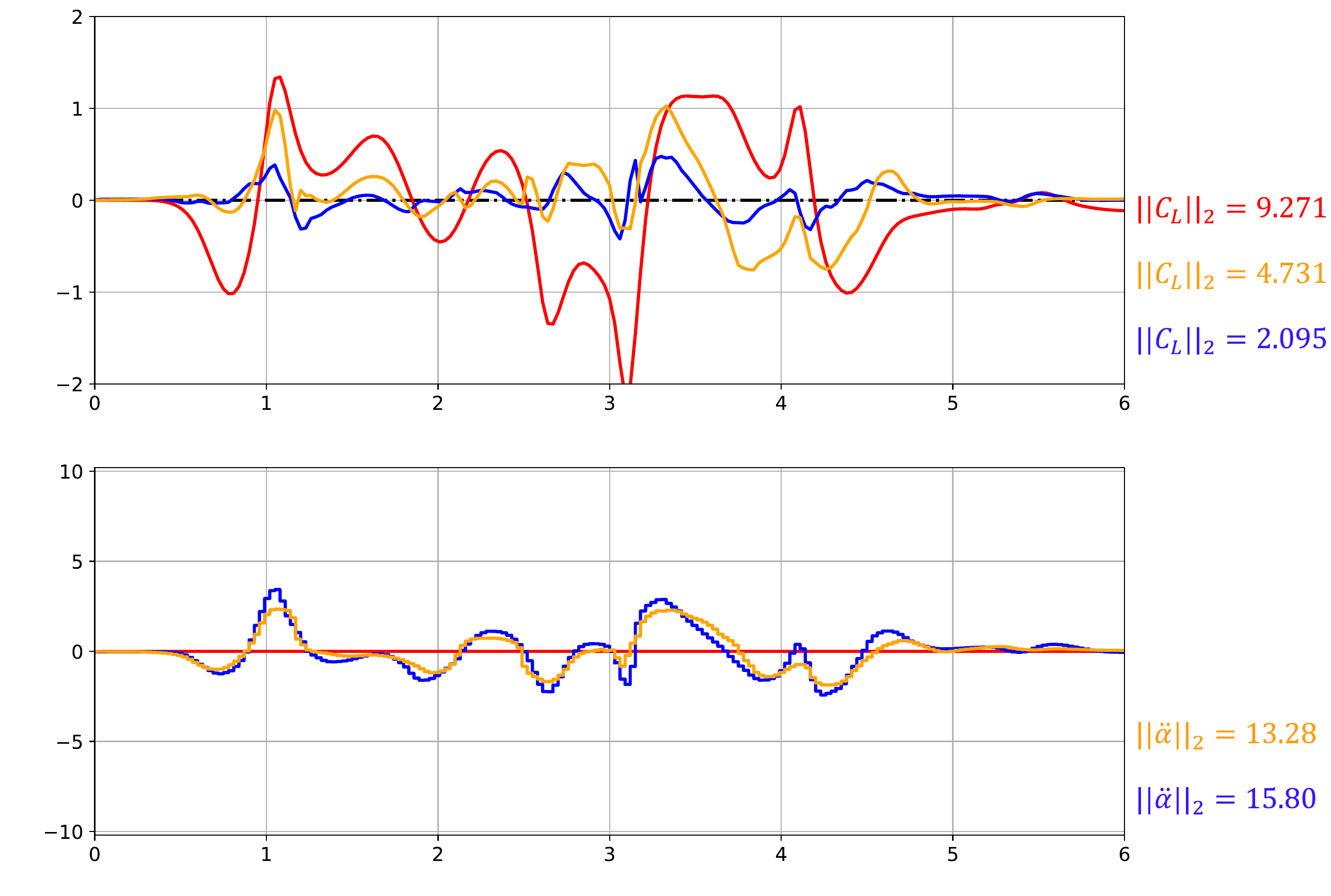}
    \put(1,50.5){\rotatebox{90}{$C_L$}}
    \put(45,34){{$t^*$}}
    \put(1,18){\rotatebox{90}{$\ddot{\alpha}$}}
    \put(45,0){{$t^*$}}
    \end{overpic}
    \caption{The lift histories and angular acceleration histories under no control, RL control trained from single-gust environment, and RL control trained from three-gust environment for one representative case in eight gust environment. Both RL controls are under quarter-chord pitching configuration. The color convention is \textcolor{red}{\rule[0.5ex]{1.5em}{1pt}: no control}, \textcolor{orange}{\rule[0.5ex]{1.5em}{1pt}: RL control trained in single-gust environment}, \textcolor{blue}{\rule[0.5ex]{1.5em}{1pt}: RL control trained in three-gust environment}. The corresponding total rewards are: \textcolor{red}{$\sum r_t=-754.95$}, \textcolor{orange}{$\sum r_t=-51.42$}, \textcolor{blue}{$\sum r_t=146.98$}.}
    \label{fig:lift_compare_eight_quarter}
\end{figure}
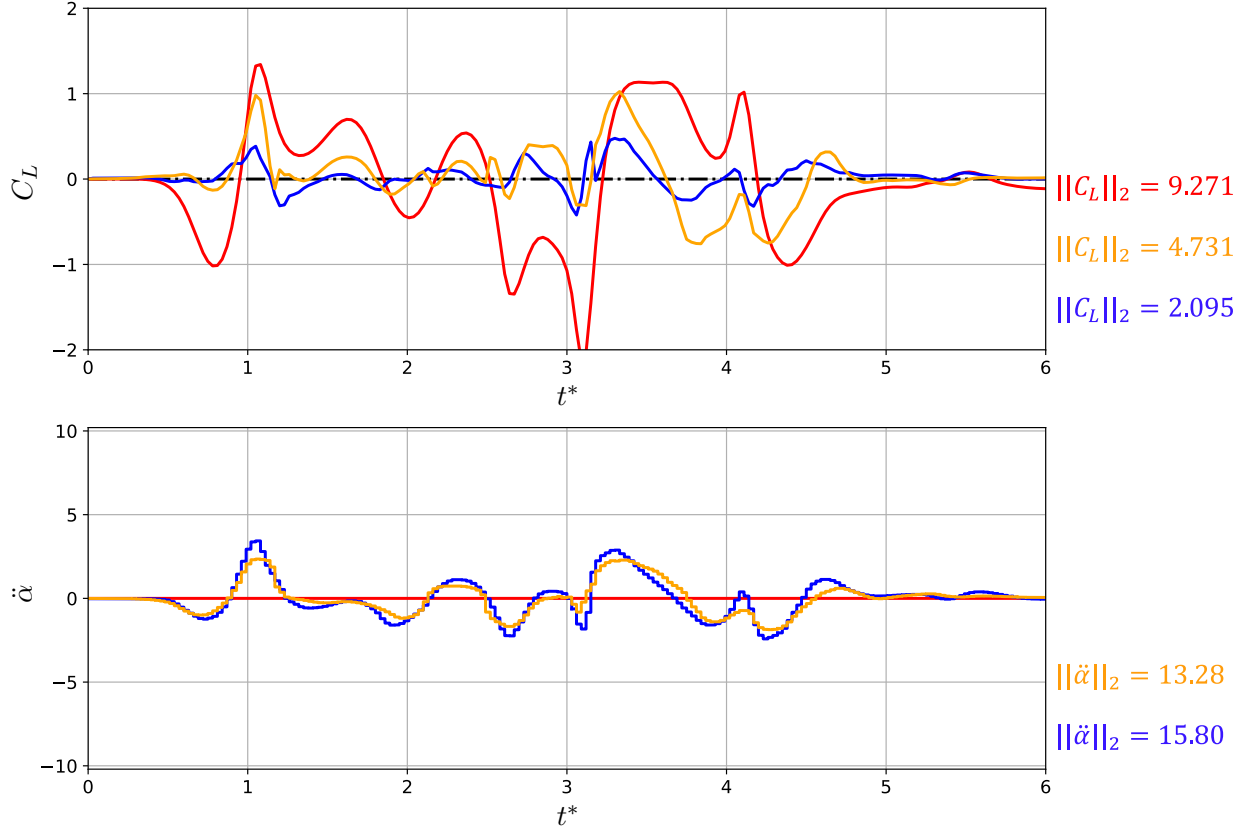

Finally, we turn our attention to the more realistic and challenging case, approximating the \textit{n}-successive gust scenario. We apply the agents trained for quarter-chord pitch control in the single- and three-gust environments, respectively, to an eight successive gust environment. As we have discussed in Section \ref{subsec:extending}, the vast parameter space of eight gust renders an exhaustive statistical evaluation infeasible, so we limit our analysis to a representative case to gain insight into the control effectiveness, as shown in Fig. \ref{fig:lift_compare_eight_quarter}. Since the simulation for this case lasts $6$ convective time units (200 control steps), the ideal total reward is $200$ when zero lift is achieved without control effort for every control step. First, the $||C_L||_2$ and $||\ddot{\alpha}||_2$ indicate that the RL agent trained in the three-gust environment can mitigate the lift variation much better than one trained in a single-gust environment, with only marginally more control effort. As a result, the total reward for the RL agent trained in the three-gust environment is much higher than the one for RL agent trained in the single-gust environment. Additionally, we find once again the familiar result that RL control in a quarter-chord pitching configuration has a higher control authority than mid-chord pitching, so that its control capability is not affected by the magnitude limit of the actuation, even in this more complex eight-gust environment. Furthermore, the RL agent trained in the three-gust environment exhibits a refined control strategy: although the angular acceleration profile shows only moderate adjustments in magnitude compared to the single-gust agent, the resulting lift response is significantly improved. This suggests that the agent has learned to leverage all lift components more effectively to compensate for interacting gust disturbances. Lastly, considering that the RL agent trained in the three-gust environment has achieved a total reward of $\sum r_t=146.98$, comparable to the ideal total reward of $200$, we can conclude that the control strategy learned in the environment involving a finite number of gusts can be effectively applied to the environment with an infinite number of gusts.

\section{Conclusions}
\label{sec:conclusions}
In this study, we have sought an effective pitch control strategy for lift regulation when a flat plate airfoil encounters an arbitrarily long sequence of gust disturbances and investigated the effect of the pivot location on the control performance and the accompanying flow response.

We have proposed a transformer-based RL framework to deal with the POMDP challenge that is common in highly disturbed flows. In light of the potentially expensive training cost of RL for such problems, we leverage two approaches that can accelerate the RL training process: pretraining and transfer learning. We have demonstrated the effectiveness of these acceleration methods in two flow control scenarios. In a mid-chord pitching configuration, we have employed P control to generate expert data for pretraining and shown that the subsequent RL training converges much faster than training from scratch. In a quarter-chord pitching configuration, we have accelerated the RL training in a three-gust environment by transferring the knowledge that the RL agent learns in a single-gust environment. \textcolor{black}{It is important to note that the gust setting in this study is markedly harder than canonical periodic flow control problems: each gust sequence produces a significantly different flow trajectory even under the same control policy. The resulting high variance makes policy and value updates prone to overfitting and to thus converge slowly. The problem is further complicated by partial observability, due to limited sparse surface pressure measurements. Our use of a transformer for belief-state inference, together with pretraining and transfer learning, has been designed to mitigate these sources of difficulty.}

We have found that P control is an effective control strategy in the mid-chord pitching configuration on its own, but that the performance improvement that RL control exhibits over P control widens for multiple-gust scenarios. In the quarter-chord pitching configuration, we have found that the RL control significantly outperforms the P control, even in a single-gust environment. Furthermore, we have shown that, in either pivot configuration, an RL control strategy learned in an environment with a finite sequence of gusts can be translated effectively into an environment with an arbitrarily long sequence of gusts.

By comparing the two pivot configurations, we have shown that the quarter-chord pitching control strategy can achieve superior lift regulation with significantly less control effort than the mid-chord pitching control strategy. Through the decomposition of the lift coefficient into $C_{L_{\amass}}$ (added-mass effect), $C_{L_{\gust}}$ (gust effect under no control), and $C_{L_{\resp}}$ (gust response under control), we have revealed that this advantage derives from the term $C_{L_{\ddot{\alpha}}}$ in the added-mass component, where it dominates the lift contribution in quarter-chord pitching but is unavailable in mid-chord pitching. In the latter configuration, the agent can only act through $C_{L_{\dot{\alpha}}}$ and $C_{L_{\resp}}$, which are comparable in magnitude to each other, but less responsive than $C_{L_{\ddot{\alpha}}}$. This difference provides more control authority in the quarter-chord pivot configuration, and thus addresses the actuation saturation problem observed in mid-chord pitching control.

\revision{The present study shows the feasibility of the transformer-based RL approach, which incorporates pretraining and transfer learning to accelerate training, for the control of a highly disturbed flow.} However, there are many ways in which the present study can be expanded in future work. For example, we do not investigate the effect of the observation history length, nor the number and locations of the pressure sensors, on the control performance. Additionally, other types of control input (e.g., synthetic jet) and/or other objectives (e.g., drag reduction, non-zero target lift) can be investigated by following the same methodology. These studies could address the underlying question of controllability, which we have not explored in this work. \revision{Furthermore, we do not explore alternative shapes of the reward function, including the relatively weighting of the lift tracking error term and the actuation penalty term, or other forms of the actuation penalization (e.g. a penalty on control effort). A systematic study of these reward-design choices would also be an interesting direction for future work. Finally, a comprehensive study of the transformer architecture (e.g. different numbers of encoder layers and attention heads) under matched conditions is another promising avenue for future work.}

\begin{acknowledgements}
    The authors gratefully acknowledge the financial support for this work provided by the National Science Foundation under Award number 2247005.
\end{acknowledgements}

\appendix

\section{\revision{A comparison of different observation window sizes}}
\label{appendix}
\revision{The choice of $N=20$ corresponds to $0.6$ convective time units, which we selected based on the flow physics in this problem setting: we found that the dominant lift variation comes from the vortex-foil interactions near the leading edge, and the boundary layer relaxes relatively quick at $Re=200$ when the gust is passing the foil. Therefore, we expect that $N=20$ already spans the main memory of the disturbance. Furthermore, this window size leads to a relatively lightweight training, which was also empirically demonstrated to enable the RL agent to learn an effective control policy across all gust scenarios considered in the present work, as we discussed in the main text.} 

\revision{To provide a direct comparison, we trained the RL agents with the observation window sizes $N=20$ and $N=40$ in a three-gust environment, as shown in figure \ref{fig:comparison_window}.
    \begin{figure}[htbp]
    \centering
    \begin{overpic}[width=1.0\linewidth]{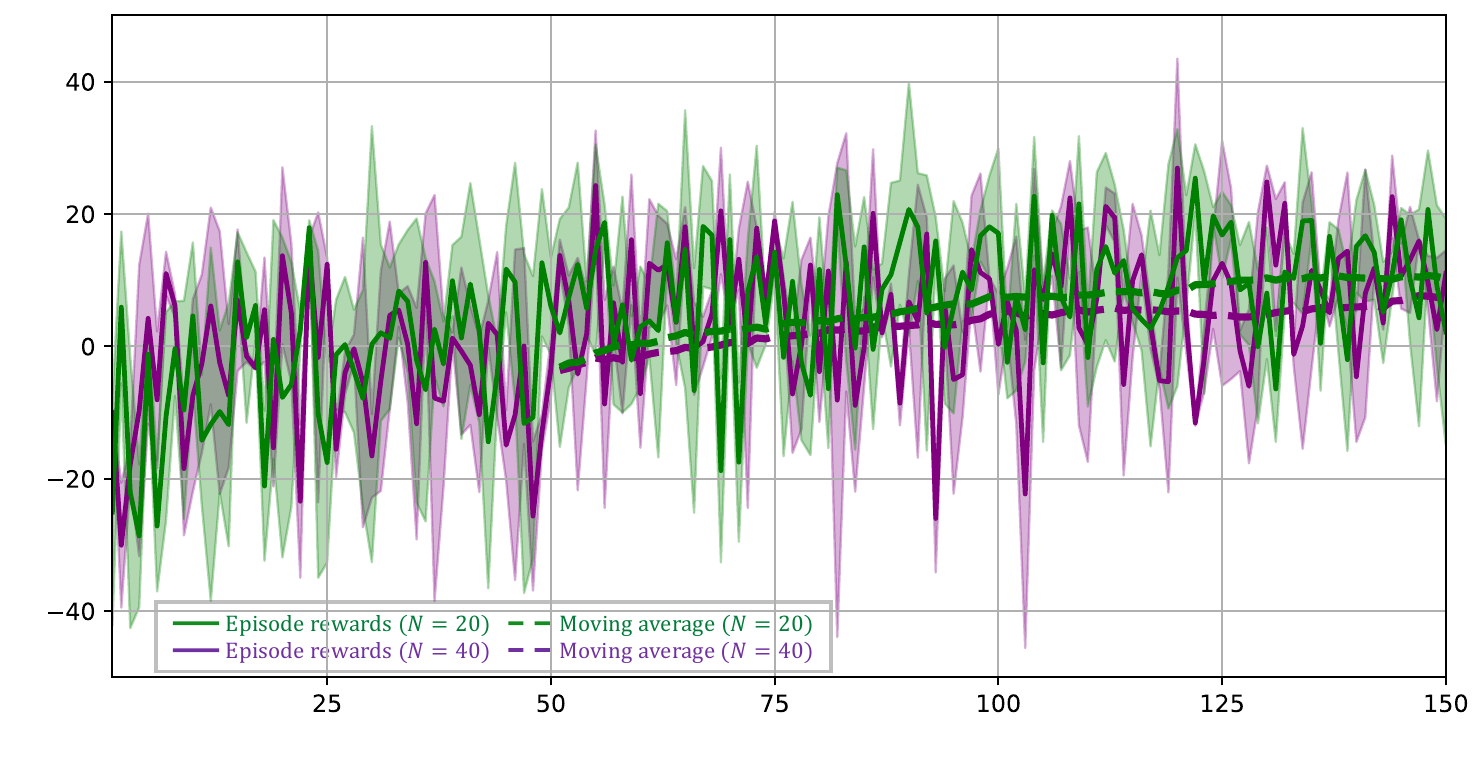}
    \put(1,19){\rotatebox{90}{Episode Rewards}}
    \put(48,1){{Episodes}}
    \end{overpic}
    \caption{\revision{A comparison of the learning curves for the RL agents having observation window size $N=20$ and $N=40$ in a three-gust environment. The solid lines and the shadow regions are the mean and std values of the episode sum rewards across three different runs. For the dashed lines, the moving average has a window size of $50$. At each episode index, the episode reward is averaged across all parallel environments per run.}}
    \label{fig:comparison_window}
    \end{figure}
    We note that the training settings are identical for the two cases (i.e. starting from the same $K_p=30$ pretrained policy, the same PPO parameters, the same training hyper-parameters, etc.). As we can observe, the $N=40$ case actually converges slightly slower than the $N=20$, and the mean episode reward for $N=40$ until $150\text{th}$ episode is also slightly lower than the $N=20$. In other words, simply doubling the observation window size without retuning the RL hyper-parameters does not necessarily improve the performance, and can even degrade it. This also corresponds to the above interpretation: once the observation window size covers most of the relevant (useful) information, adding old, weakly informative observations can possibly dilute the learning signal and thus be even harmful for the training without any retuning. However, we do not exclude the possibility that a retuned $N=40$ configuration (e.g. different learning rate, batch size, and the transformer architecture depth and width) may improve upon $N=20$, but given a higher training cost and lack of clear benefit from this experiment, we believe that the choice of $N=20$ is an efficient and sufficient choice in the present study.}

\bibliography{aps}
\end{document}